\documentclass[preprint,11pt]{elsarticle}
\usepackage[margin=3cm]{geometry} 
\usepackage{bm,upgreek}
\usepackage[utf8]{inputenc} % allow utf-8 input

\usepackage{amssymb}                              % for calligraphic letters
\DeclareMathAlphabet\mathbfcal{OMS}{cmsy}{b}{n}   % for calligraphic letters

\usepackage{lineno,hyperref}
 \usepackage{graphicx}
 \usepackage{float}
\modulolinenumbers[1]
\usepackage{amsmath}
\usepackage[export]{adjustbox}
\usepackage{subcaption}
\usepackage{longtable}
\usepackage{xcolor}
\usepackage{lineno}
\linenumbers
\usepackage{afterpage}
\usepackage{etoolbox}
\usepackage{color,soul}
\usepackage{xcolor}
\usepackage{soul}
\let\today\relax
\makeatletter
 \def\ps@pprintTitle{%
    \let\@oddhead\@empty
    \let\@evenhead\@empty
    \def\@oddfoot{\footnotesize\itshape
         { } \hfill\today}%
    \let\@evenfoot\@oddfoot
    }
\begin{document}
\nolinenumbers
\begin{frontmatter}
\title{\emph{CASU2Net}: Cascaded Unification Network by a Two-step Early Fusion for Fault Detection in Offshore Wind Turbines}

\author {Soorena~ Salari \fnref{fn1}\corref{cor1}}
\ead{soorena.salari374@gmail.com}

\author {Nasser~Sadati\fnref{fn1}}

\cortext[cor1]{Corresponding author}
\fntext[fn1]{Department of Electrical Engineering, Sharif University of Technology, Tehran, Iran}

\begin{abstract}
%% Text of abstract
This paper presents a novel feature fusion-based deep learning model (called \emph{CASU2Net}) for fault detection in offshore wind turbines. The proposed \emph{CASU2Net} model benefits of a two-step early fusion to enrich features in the final stage. Moreover, since previous studies did not consider uncertainty while model developing and also predictions, we take advantage of Monte Carlo dropout (MC dropout) to enhance the certainty of the results. To design fault detection model, we use five sensors and a sliding window to exploit the inherent temporal information contained in the raw time-series data obtained from sensors. The proposed model uses the nonlinear relationships among multiple sensor variables and the temporal dependency of each sensor on others which considerably increases the performance of fault detection model. A 10-fold cross-validation approach is used to verify the generalization of the model and evaluate the classification metrics. To evaluate the performance of the model, simulated data from a benchmark floating offshore wind turbine (FOWT) with supervisory control and data acquisition (SCADA) are used. The results illustrate that the proposed model would accurately disclose and classify more than 99\% of the faults. Moreover, it is generalizable and can be used to detect faults for different types of systems.

\end{abstract}


\end{frontmatter}
% \linenumbers
\section{Introduction}
\label{S:1}
In recent years, as a result of the global warming phenomena, wind turbines have played an important role in generating reliable, cost-effective, and pollution-free energy. Although the complex design, expensive platforms, long-distance cabling, and maintenance cost have made the electricity production cost of FOWTs three times higher than that of onshore ones, FOWTs are more advantageous owing to much steadier winds at sea, less noise pollution, and less negative impact on the environment. Moreover, the development of offshore wind energy requires careful design, planning, and installation of offshore wind farms, where their reliability can be ensured using efficient fault detection model. 
Actuator and sensor faults are common fault types that mostly occur in wind turbines. They usually arise due to mechanical and electrical reasons in most subsystems of the FOWTs, either weakly or severely. Based on two fault detection methods, i.e. model-based and data-driven methods, various techniques for fault detection in wind turbines have been proposed. Model-based methods rely on exact and highly reliable models, while wind turbines work under harsh environments and are subjected to large wind fluctuations. Therefore, we use data-driven methods, where the faulty situations can be detected in minimal time. 

Data-driven methods are increasingly used in fault detection  of wind turbines as a result of the difficulty in system modeling and, moreover, the availability of data \cite{peng2021compressive,zhang2020scada,li2020imbalance}. 

Nowadays, machine learning and deep learning methods are increasingly being used in a wide range applications such as computer vision \cite{liu2021dna, wang2021salient, pourpanah2020review, aboah2021vision}, natural language processing \cite{abdar2020energy, zhang2021building}, signal processing \cite{zhang2021hybrid, hammad2021automated}, economics\cite{magazzino2021nexus}, medical data analysis \cite{wang2020recent, abdar2021uncertainty} and many more. Unlike shallow networks, deep learning can deal with massive data sets without using feature selection, and it has had a breakthrough in certain applications \cite{goodfellow2016deep}. In wind turbines, we have big data sets that shallow learning models can not handle. So, we approached the problem of offshore wind turbine fault detection through deep learning models like long short term memory (LSTM), convolutional neural networks (CNNs), and convolutional long short term memory (ConvLSTM). The authors in \cite{helbing2018deep} discussed the application of various deep learning models for fault detection in wind turbines. In \cite{lei2019fault}, a fault detection and classification model using long short-term memory (LSTM) employing sensor data generated from a real wind turbine is presented. In another study \cite{jiang2018multiscale}, Jiang et al. used a multiscale CNN for fault detection in wind turbines. They utilized a non-overlapping sliding window to process the data and then employed multiscale CNN for feature extraction and classification. In \cite{qiao2015survey1} and \cite{qiao2015survey2}, the principal information on fault diagnosis and condition monitoring of wind turbines have been investigated. A comprehensive review in wind turbine condition monitoring is also given in \cite{hameed2009condition}. A fault detection model based on image texture analysis using statistical machine learning methods such as K nearest neighbors (KNN), bagged trees, and decision tree is presented in \cite{ruiz2018wind} with more than 80\% accuracy. A sliding window in the data preprocessing and denoising autoencoder with PCA for fault detection of wind turbines is suggested in \cite{jiang2017wind}. A fault detection model using a random forest classifier with actuator and sensor data for offshore wind turbines is also presented in \cite{si2017data}. An ensemble of some statistical models like support vector machine (SVM) and multi-layer perceptron (MLP) also is used for fault detection in offshore wind turbines \cite{pashazadeh2018data}. 

Information fusion originates from data fusion and has various applications like medical image and signal fusion \cite{yao2020multi,he2020feasibility}, sensor data fusion \cite{shanshanoriginal}, and feature fusion \cite{xie2018fusing}. The authors in \cite{hang2014fault} used some methods like fuzzy integral, D-S evidential reasoning, and ordered weighted averaging to fuse the fault detection results. In another study \cite{peng2021wind}, a novel feature fusion model was used for the concatenation of information from gearbox vibration and generator current signals. They created two independent classifiers with probabilistic output and combined the outputs based on the Dempster–Shafer theory and the softmax regression technique. Xiao et al. \cite{xiao2021misalignment} extracted features in time-domain, frequency domain, and time-frequency domain from several sensors like vibration, stator current, and temperature. After that, they employed a least square support vector machine (LSSVM) to calculate the probability of the classes and Dempster–Shafer theory for information fusion and fault diagnosis based on the outputs of LSSVM.

Undoubtedly, we confront uncertainty in real-world applications like weather forecasting, image classification, image segmentation, medical signal processing, and many more \cite{abdar2021uncertainty}. So, it is very crucial to estimate the trustability of machine learning (ML) and deep learning (DL) models \cite{abdar2021review}. The predictions of such models are not reliable and can take expensive tolls on the system in some areas like wind turbines. In this regard, uncertainty quantification (UQ) methods have a vital role in uncertainty reduction and boosting the trustworthiness of the models. This allows ML and DL models to learn how deal with unknown samples. In other words, these models can be ready to say \textcolor{red}{\emph{I DO NOT KNOW}} or \textcolor{red}{\emph{I AM NOT SURE}} \cite{abdar2021review, abdar2021uncertainty, abdar2021barf} when they are unsure of their predictions.
%\subsection{Information fusion} 

%\subsection{Uncertainty quantification (UQ)}

This paper presents a novel and significantly efficient deep learning model with two steps feature fusion of multiple sources. These feature fusion strategies helped to have both high-level and low-level features and boost the fault detection performance. Plus, we considered Ensemble MC dropout \cite{gal2016dropout} to consider uncertainty in the model and have a trustable model. Also, we visualized the features of the discussed models with T-SNE \cite{van2008visualizing} to have a detailed understanding of the models' operation. The main features of the proposed method can be summarized as follows:
\begin{enumerate}

\item	We proposed a novel model with ConvLSTM layers as feature extractors, two information fusion steps from different layers of the networks, and uncertainty quantification for fault detection in offshore wind turbines called \emph{CASU2Net}. \emph{CASU2Net} also considers nonlinear correlation among multivariate sensor data. To the best of our knowledge, this paper is the first to use uncertainty quantification combined with information fusion for fault detection in offshore wind turbines.

\item	A sliding window is considered for inherent time dependency present in the data set. Thus, the proposed models use new data that contains current and past relationships among sensors, which increases the accuracy of fault detection.

\item	Sensors are costly components in offshore wind turbines. Therefore, unlike other fault detection models that use 13 and 8 sensors for fault detection in offshore wind turbines \cite{ruiz2018wind}\cite{lei2019fault}, we only use 5 sensors and improve the fault detection accuracy.
\end{enumerate}

The data set is derived from the SCADA developed by the National Renewable Laboratory (NREL) for fault detection and fault-tolerant control of a benchmark MW-scale wind turbine  \cite{odgaard2013fault_TCST,odgaard2013wind_ACC}. This benchmark provides the same model for all researchers to evaluate their fault detection models in the presence of common faults. Thus, after gathering data and preparing labels, we preprocess the data. Then, we train and test the models with 10-fold cross-validation. Finally, simulation results are  evaluated by various machine learning metrics to show their performance in classifying faulty cases from the normal ones.

%The rest of this research is structured as follows. In Section \ref{S:2}, we discuss the benchmark model used for simulating and collecting data. Section \ref{S:3} explains the application of deep learning methods for fault detection.

\section{Benchmark Model}
\label{S:2}

In this section, we introduce the benchmark model used for simulating and collecting data, as proposed in \cite{odgaard2013wind_ACC}. This benchmark model is a 3-bladed horizontal axis variable speed wind turbine consisting of a fully coupled converter, which its rated power is 5 MW. The aim of this benchmark model is to provide a base model with which all researchers can compare their results. This benchmark is based on software developed by NREL called fatigue, aerodynamics, structures, and turbulence (FAST), which is an aeroelastic wind turbine simulator \cite{jonkman2005fast}. With this powerful software, researchers can quickly simulate the wind turbine structural behavior with up to 24 degrees of freedom. In this paper, FAST numerical simulations of the NREL 5-MW offshore wind turbine are accomplished. The characteristics of this turbine can be found in Table \ref{tab:1} \cite{jonkman2009definition}. Wind data sets are also generated by TurbSim \cite{jonkman2012turbsim}, to make the methodology more realistic. The wind has the following characteristics:
\begin{enumerate}
  \item Kaimal turbulence model with $10 \%$ intensity,
  \item Mean speed of 18.2 $\mathrm{~m} / \mathrm{s}$ at the hub height.
\end{enumerate}

\begin{figure}[h]
\centering
            \includegraphics[width=1\textwidth]{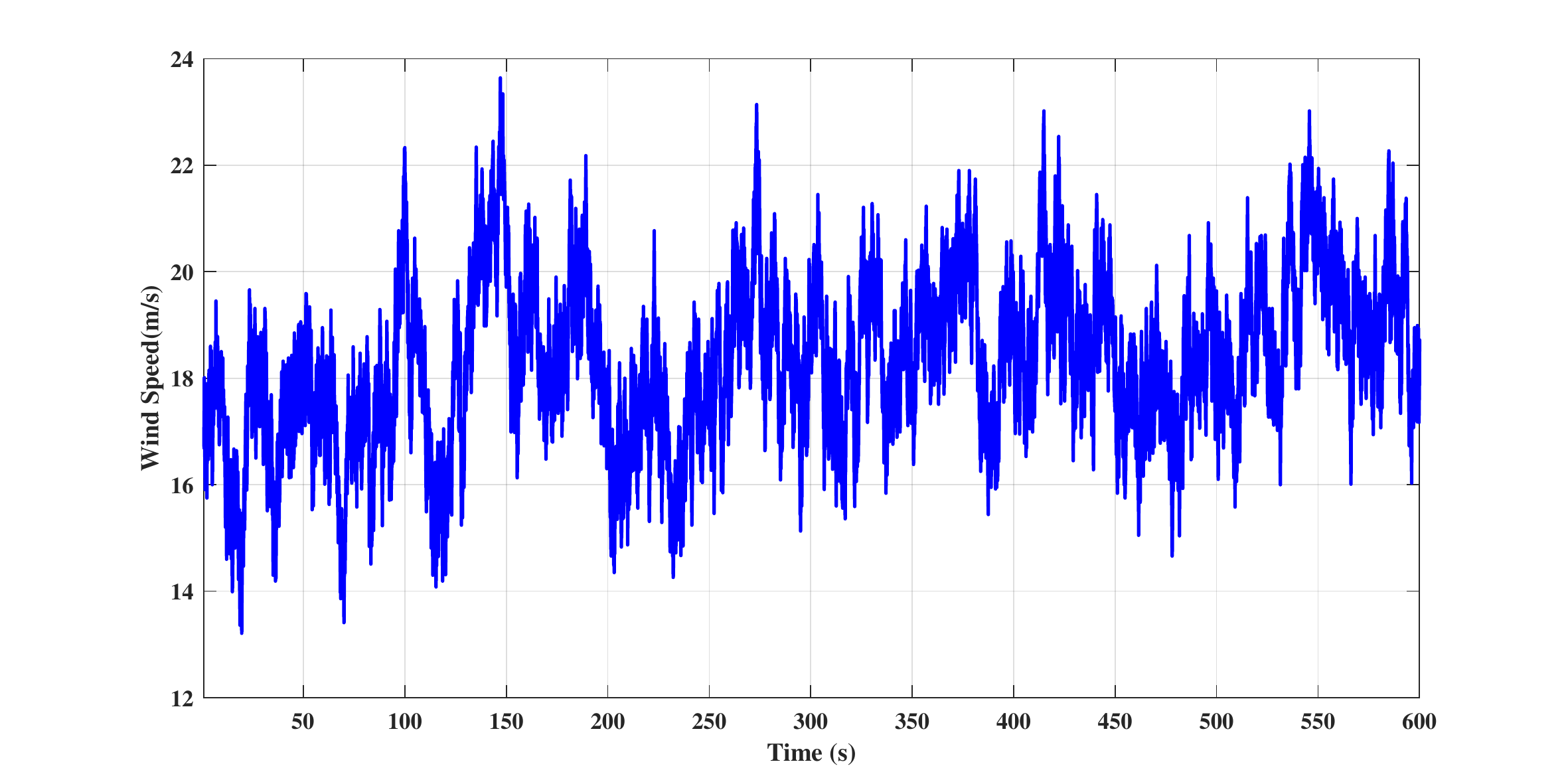}
     %\vspace*{0.1cm}
    \caption{A generated wind profile.}
    \label{wind}
\end{figure}

Fig. \ref{wind} shows sample of a generated wind profile with duration of 600 s.

\begin{table}[h]
\caption{The NREL 5-MW wind turbine characteristics [13].}
\centering
\begin{tabular}{l l}
\hline
Property & Value \\
\hline
Rated power  & $5 MW$ \\
Number of blades& $3$\\
Height of tower & $87.6 m$ \\
Gearbox ratio & $98$ \\
Rotor diameter & $126 m$ \\
Cut-in, rated, cut-out wind speed & $3 m/s, 11.4 m/s, 25 m/s$ \\
Nominal generator speed & $1173.7 rpm$ \\
\hline
\end{tabular}
\label{tab:1}
\end{table}

\subsection{Generator - converter actuator model}
The dynamic equation of a generator coupled with a converter is as follows \cite{odgaard2013wind_ACC}:

\begin{equation}
\label{eq1}
\dot{\tau}_{g}+\alpha_{g} \tau_{g}(t)=a_{g c} \hat{\tau}_{g r}(t)
\end{equation}
where $\hat{\tau}_{g r}$ is the generator torque reference signal, $\tau_{g}$ is the generator torque, and $\alpha_{g c}$ is the converter model parameter. Also, the nominal value of $\alpha_{g c}$ is 50. The generator output power is given by:

\begin{equation}
\label{eq2}
P_{g}(t)=\eta_{g} \omega_{g}(t) \tau_{g}(t)
\end{equation}

where the nominal value of $\eta_{g}$ is 0.98.

\subsection{Generator - converter actuator model}
The dynamics of a hydraulic pitch angle actuator is as follows:

\begin{equation}
\label{eq3}
\ddot{\beta}+2 \zeta \omega_{n} \dot{\beta}+\omega_{n}^{2} \beta=\omega_{n}^{2} \hat{\beta}_{c}
\end{equation}

where $\omega_{n}$ is the natural frequency and $\zeta$ is the damping coefficient. Equation (3) corresponds to all three blades. The nominal values of $\zeta$ and $\omega_{n}$ are  0.7 and 11.11 rad/s, respectively \cite{odgaard2013wind_ACC}.

\subsection{Baseline controller}

The 5-MW reference wind turbine has both collective pitch controller and torque controller, where their input is generator's speed. These baseline controllers are commonly used among researchers for the purpose of comparison \cite{ochs2013simulation, vidal2012power, beltran2008sliding}. To reduce the high-frequency dynamics of the control system in the gain-scheduling PI (GSPI) controller, a low-pass filter is used for generator speed \cite{jonkman2009definition}. 
The baseline torque controller is given by:				\begin{equation}
\label{eq4}
\tau_{g, r}(t)=\frac{P_{r e f}(t)}{\hat{\omega}_{g}(t)}
\end{equation}		
where $\hat{\omega}_{g}(t)$ is the output of the low-pass filter, and $P_{r e f}(t)$  represents the reference power.

A baseline collective pitch controller is used as a GSPI controller with generator speed as its input \cite{jonkman2009definition}. The nonlinearities of the wind turbine are compensated by changing the controller's gains. The controller is given by:

\begin{equation}
\label{eq5}
\beta_{c}=K_{p}(t)\left(\hat{\omega}_{g}(t)-\omega_{g r}(t)\right)+K_{i}(t) \int_{0}^{t}\left(\hat{\omega}_{g}(t)-\omega_{g r}(t)\right)
\end{equation}
where $\omega_{g r}$ is the generator speed reference, and $\beta_{c}$ represents the pitch angle set point.

\subsection{Data set}

In this section, we present in details the simulated data set used in this work for training and evaluating the obtained model. All the sensors used in the benchmark model appear in Table \ref{tab:50} \cite{odgaard2013wind_ACC}. These sensors, which cover every part of a turbine, are typically used in the MW-scale commercial wind turbines.

In the fault detection phase, 140 simulations for healthy cases and 40 simulations for each fault have been carried out. The sampling frequency is 80 Hz in each simulation. Therefore, we have 420 cases with 13 variables in our data set.

The more sensors deployed in a given system, the more we need to spend; hence, reducing the number of sensors is necessary. We use backward feature selection \cite{borboudakis2019forward} to find out the importance of each feature. After observing the impact of eliminating each sensor, we have selected only 5 sensors out of 13 sensors and use them in our reduced data set. Our selected sensors are: generator torque, rotor speed, and all three pitch angles.

\begin{table}[h]
%\hspace{-30cm}
\caption{Available sensors on a MW-scale industrial turbine \cite{odgaard2013wind_ACC}.}
\centering
\begin{tabular}{l l l l}
\hline
Number & Sensor type & Units \\
\hline
\hline
1 & Rotor speed & rad/s\\
2 & Generator torque & Nm\\
3 & First pitch angle & deg\\
4 & Second pitch angle & deg\\
5 & Third pitch angle & deg\\
6 & Electrical power & $kW$\\
7 & Generator speed & rad\\
8 & Acceleration of fore-aft movement at the tower bottom & $\frac{m}{s^{2}}$\\
9 & Acceleration of side-to-side movement at the tower bottom & $\frac{m}{s^{2}}$\\
10 & Acceleration of fore-aft movement at the mid-tower & $\frac{m}{s^{2}}$\\
11 & Acceleration of side-to-side movement at the mid-tower & $\frac{m}{s^{2}}$\\
12 & Acceleration of fore-aft movement at the top of the tower & $\frac{m}{s^{2}}$\\
13 & Acceleration of side-to-side movement at the top of the tower & $\frac{m}{s^{2}}$\\
\hline
\end{tabular}
\label{tab:50}
\end{table}

\subsection{Fault scenarios}

Wind turbines are subjected to different types of fault conditions. Inspired from \cite{odgaard2013fault_TCST}\cite{odgaard2013wind_ACC}, the most often occurring faults in a wind turbine are tabulated in Table \ref{tab:fault}.

As mentioned earlier, we use both sensor faults and actuator faults. It should be noted that, since the pitch and torque actuators play a vital role in wind turbine operation, and their failure can cause the whole system to fail, we consider these actuators. In Table \ref{tab:fault}, faults 1, 2, and 3 are the pitch actuator's faults which occur due to hydraulic leakage, pump wear, and high air content \cite{chaaban2014structural}. On the other hand, as stated in \cite{liniger2015reliable} and \cite{chen2013active}, some sensor faults are more common and important compared to other sensor faults. For this reason, we choose faults 4, 5 and 6 in our case study. These latter ones could be electrical or mechanical and appear to be due to wrong gain and sensor saturation. Faults 5 and 6 could affect the baseline controller, and the pitch position. Finally, the torque actuator fault (fault 7) could appear because of wrong initialization of the converter or manufacturing failures.

\begin{table}[h]
%\hspace{-30cm}
\caption{Fault scenarios.}
\centering
\begin{tabular}{l l l l}
\hline
Fault & Type & Description  \\
\hline
\hline
1 & Pitch actuator & High air content ($\zeta =0.45$  and $\omega_{n}= 5.73$   rad/s)\\
2 & Pitch actuator & Pump wear parameters ($\zeta =0.75$ and $\omega_{n}= 7.27$ rad/s)\\
3 & Pitch actuator & Hydraulic leakage ($\zeta =0.9$ and $\omega_{n}= 3.42$ rad/s)\\
4 & Generator speed sensor & Gain factor (1.2)\\
5 & Pitch angle sensor & Fixed value (10 deg)\\
6 & Pitch angle sensor & Fixed value (5 deg)\\
7 & Torque actuator & Offset value (2000 Nm)\\

\hline
\end{tabular}
\label{tab:fault}
\end{table}

\section{Fault Detection with Deep Learning}
\label{S:3}
 
In this paper, we propose three different deep learning models, including simple CNN, multi-headed model, and \emph{CASU2Net}  to detect faults in offshore wind turbines based on deep learning. In the following, we present new deep learning approaches.

\subsection{Recurrent neural networks}
 
A recurrent neural network (RNN) is a class of neural networks that has feedback and hidden states. Recurrent neural networks have a wide variety of applications from image processing to speech and natural language processing \cite{karpathy2016convolutional}. The output of a hidden state at a specific time depends on previous times and it is given by times and it is given by
\begin{equation}
\label{eq6}
h(t)=f\left(W_{h} h(t-1)+W_{x h} x(t)\right),
\end{equation}
where $W_{h h}$ and $W_{x h}$ are the learnable parameters, $x(t)$ is the input at time instant $\mathrm{t}, h(t-1)$ is a hidden state at time $\mathrm{t}-1$, and $f$ is an activation function like tanh. The output of an RNN is given by: 

\begin{equation}
\label{eq7}
y=W_{x h} h(t)
\end{equation}
 
Similar to ordinary neural networks, recurrent neural networks suffer from vanishing and exploding gradients. To update each of the weights in the  backpropagation algorithm, we use their gradients. To compute the gradients, the chain rule is employed as follows \cite{bishop2006pattern}:
 
\begin{equation}
\label{eq8}
\frac{\partial L_{j}}{\partial W}=\sum_{k=1}^{j} \frac{\partial L_{j}}{\partial h_{k}} \frac{\partial h_{k}}{\partial W}
\end{equation} 
where W represents the network weight matrix, L represents the loss function, and $k$ is the layer number. We also use the chain rule to compute $\partial L_{j} / \partial h_{k}$, and then rewrite \ref{eq8} as follows: 
 
\begin{equation}
\label{eq9}
\frac{\partial L_{j}}{\partial W}=\sum_{i=1}^{j} \frac{\partial L_{j}}{\partial y_{j}} \frac{\partial y_{j}}{\partial h_{j}} \frac{\partial h_{j}}{\partial h_{k}} \frac{\partial h_{k}}{\partial W}
\end{equation} 

Because each state is related to previous states involving nonlinear equations, we calculate $\partial h_{j} / \partial h_{k}$ as follows: 
 
\begin{equation}
\label{eq10}
\frac{\partial h_{j}}{\partial h_{k}}=\prod_{m=k+1}^{j} \frac{\partial h_{m}}{\partial h_{m-1}}
\end{equation}
Now by combining \ref{eq9} and \ref{eq10}, we obtain:
\begin{equation}
\label{eq11}
\frac{\partial L_{j}}{\partial W}=\sum_{k=1}^{j} \frac{\partial L_{j}}{\partial y_{j}} \frac{\partial y_{j}}{\partial h_{j}}\left(\prod_{m=k+1}^{j} \frac{\partial h_{m}}{\partial h_{m-1}}\right) \frac{\partial h_{k}}{\partial W}
\end{equation}

The nonlinear function for each state of the equation, which is related to previous states, is usually a tanh or ReLU activation function, and based on \ref{eq10}, the nonlinear functions are multiplied together. If the gradient is less than 1, after many multiplications in \ref{eq10}, based on \ref{eq11} the value of gradient   function approaches zero, so the gradient and vanishes, which in turn prevents the weights from updating their values. Furthermore, if the gradient is more than 1, after many multiplications in \ref{eq10}, based on \ref{eq11} the value of gradient will become unlimited, and gradients explodes. We should remark that such a gradient computation requires a lot of memory \cite{karpathy2016convolutional}.

\subsection{Convolutional neural networks}

Convolutional neural networks (CNNs) makes a noteworthy improvement in image processing, speech processing, and neuroscience. This progress is caused by a mixture of algorithmic developments and accessing to large amounts of data and computer resources \cite{asawa2020cs231n}. CNN has two main layers: the convolutional layer and the pooling layer. The first layer is the core of the network, where the convolution operation is performed. The convolution operation is calculated as follows:

\begin{equation}
\label{eq12}
s(t)=\int x(a) w(t-a) \mathrm{da}
\end{equation}
where $x, w$ and $s$ represent the input, kernel, and the output, respectively. In convolutional network terminology, the output is known as feature map. Kernels are trainable weights and optimized during the training stage with backpropagation. The second layer is the pooling layer, which reduces the size of the feature map. As a result of this size reduction, in the next convolutional layer of the network, the kernels with lower size are needed and consequently the number of parameters for training and the amount of computation in the network are decreased. After convolution layer, the pooling layer is used for where x, w and s represent the input, kernel, and the output, respectively. In convolutional network terminology, the output is known as feature map. Kernels are trainable weights and optimized during the training stage with backpropagation. The second layer is the pooling layer, which reduces the size of the feature map. As a result of this size reduction, in the next convolutional layer of the network, the kernels with lower size are needed and consequently the number of parameters for training and the amount of computation in the network are decreased. After convolution layer, the pooling layer is used for taking subsamples from the feature map, which has been produced by the convolutional layer. The subsample operation usually takes the maximum or average of the feature map. This layer helps the network to recognize patterns in data set and makes the network more robust against rotation and scaling. By these two layers, CNN uses local features and nonlinear transformation, and it can also extract distinctive patterns by learning from data.

\subsection{Long short-term memory}
The LSTM is a kind of recurrent neural network that relates to time series, and it can solve the vanishing gradient as well as exploding gradient problem \cite{hochreiter1997long}. Fig. \ref{LSTM_structure} shows the LSTM structure.

The LSTM equations are given as follows:

\begin{figure}[h]
\centering
            \includegraphics[width=1\textwidth]{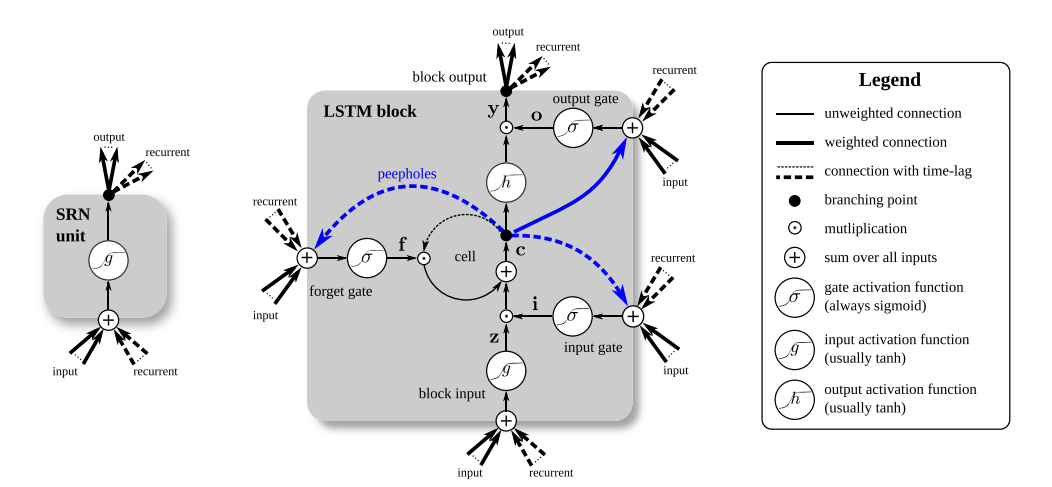}
     %\vspace*{0.1cm}
    \caption{LSTM structure \cite{greff2016lstm}.}
    \label{LSTM_structure}
\end{figure}

\begin{equation}
\label{eq13}
z_{t}=g\left(W_{z} x_{t}+R_{z} y_{t-1}+b_{z}\right)
\end{equation}

\begin{equation}
\label{eq14}
i_{t}=\sigma\left(W_{i} x_{t}+R_{i} y_{t-1}+p_{i} \odot c_{t-1}+b_{i}\right)
\end{equation}

\begin{equation}
\label{eq15}
c_{t}=z_{t} \odot i_{t}+c_{t-1} \odot f_{t}
\end{equation}

\begin{equation}
\label{eq16}
f_{t}=\sigma\left(W_{f} x_{t}+R_{f} y_{t-1}+p_{f} \odot c_{t-1}+b_{f}\right)
\end{equation}

\begin{equation}
\label{eq17}
y_{t}=h\left(c_{t}\right) \odot o_{t}
\end{equation}

\begin{equation}
\label{eq18}
o_{t}=\sigma\left(W_{o} x_{t}+R_{o} y_{t-1}+p_{o} \odot c_{t}+b_{o}\right)
\end{equation}
where $z_{t}$ is the input block, $i_{t}$ is the input gate, $c_{t}$ is the memory cell, $f_{t}$ is the forget gate, $y_{t}$ is the output and $O_{t}$ is the output gate. Also $W_{z}, W_{i}, W_{f}$, and $W_{o}$ denote the input weight matrices for the input gate, forget gate and the output gate; $p_{i}, p_{f}$, and $p_{o}$ are the peephole weight matrices; $R_{z}, R_{i}, R_{f}$, and $R_{o}$ are recurrent weight matrices; $b_{z}, b_{i}, b_{f}$, and $b_{o}$ are bias weight matrices; and the operator $\bigodot$ means element-wise multiplication. In addition, $\sigma, g$ and $h$ represent the logistic functions, usually considered as sigmoid (see \ref{sigmoid}) or tanh (see \ref{tanh}) activation functions. $Sigomid$ maps the inputs between 0 and 1, while $tanh$ is a hyperbolic function that have outpup between -1 and 1.

\begin{equation}
 sigmoid (x)=\frac{1}{1+e^{-x}}
\label{sigmoid}
\end{equation}

\begin{equation}
\tanh (x)=\frac{e^{x}-e^{-x}}{e^{x}+e^{-x}}
\label{tanh}
\end{equation}

The memory cell is an important part of an LSTM network and can replace matrix multiplication with a first-order recursive equation. Therefore, vanishing and exploding gradients are eliminated and the model can be trained well \cite{greff2016lstm}.

\subsection{ConvLSTM}

ConvLSTM replaces matrix multiplication with convolution operation at each gate in the LSTM cell \cite{shi2015convolutional}. Unlike recurrent neural networks, LSTM learns long-term dependencies by assigning the forget gate to 1. In addition, when the forget gate is set to 1 during backpropagation, the derivative is always 1 and the gradient of the memory cell does not vanish. In this paper, we use ConvLSTM to benefit its power in confronting the sequential data. The ConvLSTM equations are given as follows:

\begin{equation}
\label{eq13}
z_{t}=g\left(\operatorname{Conv}(W_{z}, x_{t})+\operatorname{Conv}(R_{z},y_{t-1})+b_{z}\right)
\end{equation}

\begin{equation}
\label{eq14}
i_{t}=\sigma\left(\operatorname{Conv}(W_{i},x_{t})+\operatorname{Conv}(R_{i},y_{t-1})+\operatorname{Conv}(p_{i},c_{t-1})+b_{i}\right)
\end{equation}

\begin{equation}
\label{eq15}
c_{t}=z_{t} \odot i_{t}+c_{t-1} \odot f_{t}
\end{equation}

\begin{equation}
\label{eq16}
f_{t}=\sigma\left(\operatorname{Conv}(W_{f},x_{t})+\operatorname{Conv}(R_{f},y_{t-1})+\operatorname{Conv}(p_{f},c_{t-1})+b_{f}\right)
\end{equation}

\begin{equation}
\label{eq17}
y_{t}=h\left(c_{t}\right) \odot o_{t}
\end{equation}

\begin{equation}
\label{eq18}
o_{t}=\sigma\left(\operatorname{Conv}(W_{o},x_{t})+\operatorname{Conv}(R_{o},y_{t-1})+p_{o} \odot c_{t}+b_{o}\right)
\end{equation}

\subsection{Monte carlo dropout}

The authors in \cite{gal2016dropout} showed that simple dropout could represent a Bayesian approximation of a gaussian process. They manifested that even in predictions with high softmax outputs, there are false predictions. In this regard, we trained several models with different neurons using simple dropout and consider the networks as Monte Carlo samples from the complete set of models. This Monte Carlo sampling provides us with a mathematical basis for the evaluation of the model's uncertainty. We use MC dropout in both training time and test time. Then, we will have multiple predicted probabilities rather than one predicted probability and average them to analyze the obtained predictions. This strategy helps us to have an ensemble model of several models. You can see the MC dropout equations below:

\begin{align}
\label{eq:MC_dropout}
\hat{\mathbf{Y}}_{predict} =  \operatorname{Softmax}\left(\mathbf{\textit{Model}}\left(\mathbf{Input}; \mathbf{W}\right)\right),
\end{align}

\begin{align}
\label{eq:EMC_dropout}
\hat{\mathbf{Y}}_{output}=\frac{1}{K} \sum_{0}^{K} \hat{\mathbf{Y}}_{predict}.
\end{align}

\begin{align}
\label{eq:SOF3}
Predicted\textunderscore Class =\operatorname{Argmax}(\hat{\mathbf{Y}}_{output}).
\end{align}
where $W$ and $Input$ are the weights and input of the model. We should note that we have used $K$ equal to 200 in our Monte Carlo simulations.
\subsection{Data preprocessing}

One of the challenging problems in fault detection is reducing the time needed for fault detection and accommodation. The offshore wind turbines are exposed to difficult phenomena such as wind and waves, and therefore, immediate fault detection is essential for damage prevention.

In this paper, we use an overlapping sliding window with a length of only 200 samples (2.5 s), where each sample is a 2-D matrix whose first axis is time and the second axis is the sensor measurement. In the previous work, the whole signal length of 600 s is considered \cite{ruiz2018wind}.

\subsection{Network Architecture}

\begin{figure}[h]
\centering
            \includegraphics[width=0.7\textwidth]{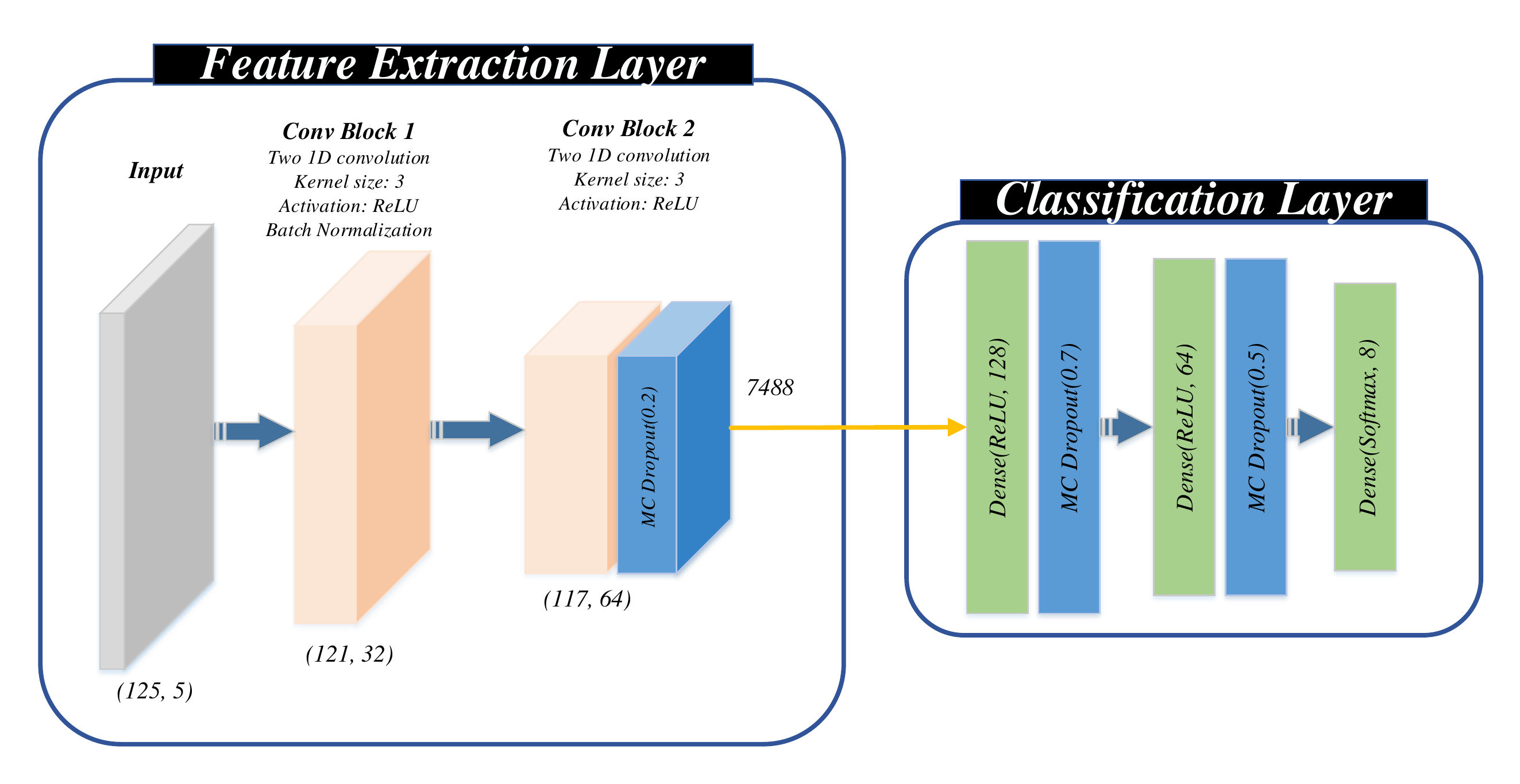}
     %\vspace*{0.1cm}
    \caption{A general overview of the applied deep learning 1 (Simple CNN).}\label{SimpelCNN_Model}
\end{figure}

Our first proposed model is constructed by CNN and called simple CNN, as shown in Fig. \ref{SimpelCNN_Model}. This model consists of the following components: 8 hidden layers (convolution and dense layers), where each convolution and dense layer is followed by an activation function. The first two convolutional layers contain 32 filters, kernel size of 3, and ReLU activation functions. Thus, a batch normalization layer and 2 convolutional layers with 64 filters and kernel size of 3 are added. After convolutional layers, MC dropout only keeps $20 \%$ of the nodes for training. Finally, the model has followed by classification layers, which consist of a batch normalization layer (this layer reduces the overfitting problem and increases the network generalization \cite{ioffe2015batch}), a dense layer with 128 hidden units and ReLU as activation function, MC dropout with a rate equal to 0.5, and a dense layer with 64 hidden unit and ReLU as activation function, and an output node connected to the dense layers with eight hidden units and softmax activation function for the classification task. In addition, model training is performed by 50 epochs, a batch size of 32, and an Adam optimizer \cite{kingma2014adam}.

\begin{figure}[h]
\centering
            \includegraphics[width=0.7\textwidth]{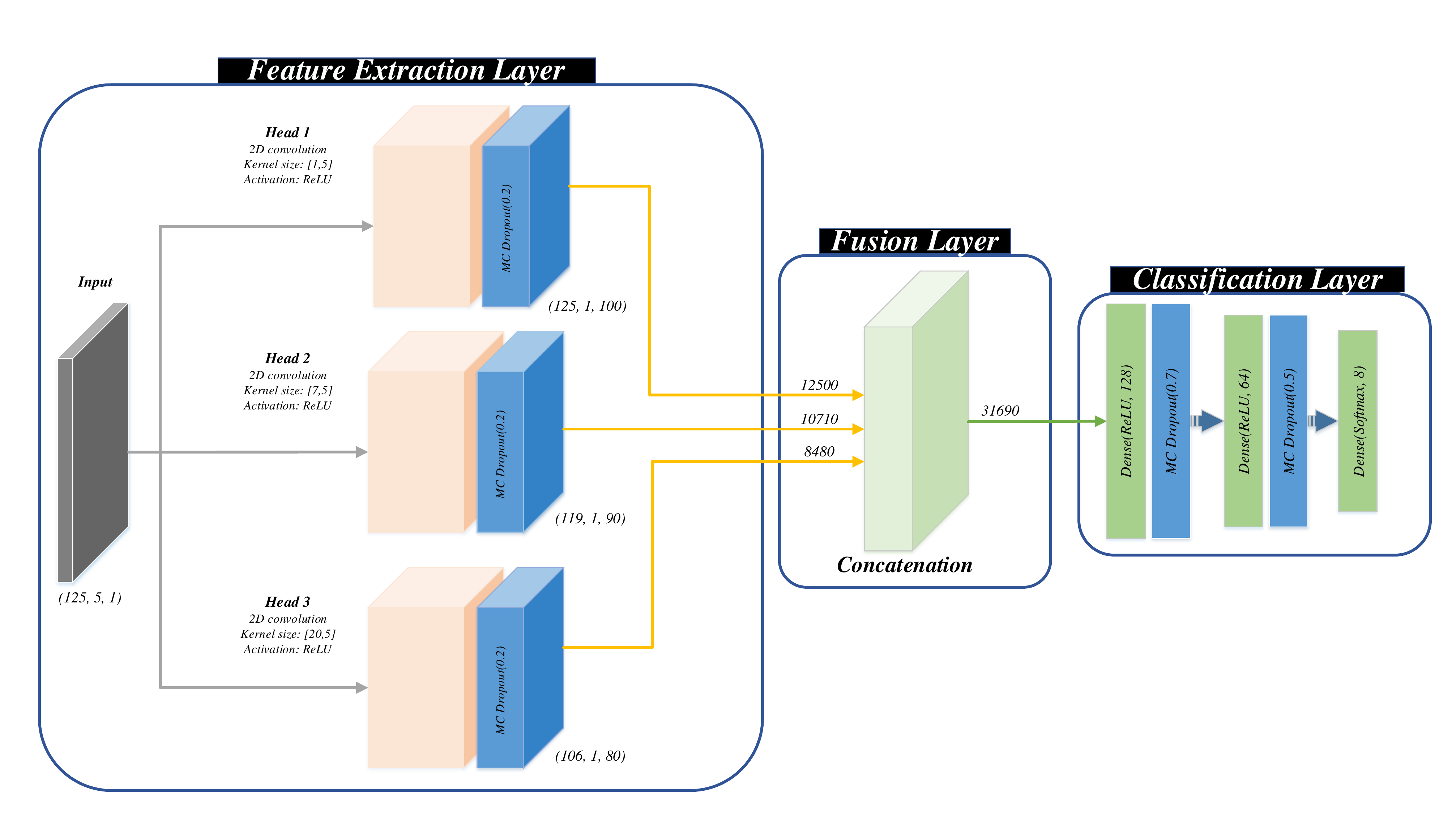}
     %\vspace*{0.1cm}
    \caption{A general overview of the applied deep 2 (Multi-headed).}\label{Multi-headed_Model}
\end{figure}

The second proposed model called the multi-headed model and constructed by CNN, is shown in Fig. \ref{Multi-headed_Model}. It consists of 3 forward ways. In each part of the network, we have a convolutional layer with 100, 90, and 80 filters with kernel sizes of $1 \times 5,7 \times 5$, and $20 \times 5$, respectively, and MC dropout with a rate equal to 0.2. The output of all convolutional layers is followed by a ReLU activation function and a batch normalization regularizer. We use concatenation of 3 convolution layers that are followed by classification layers like simple CNN, where each layer is used for a specific purpose. For example, when we use a filter with kernel size of $20 \times 5$, the relevance of each of 20 sample times with 20 nearby sample times is calculated. Because we use the concatenation of these outputs, our model is an example of the ensemble learning algorithm of convolutional layers. The model training is performed by 50 epochs, batch size of 32, and an Adam optimizer.

\begin{figure*}[h]
\centering
            \includegraphics[width=1\textwidth]{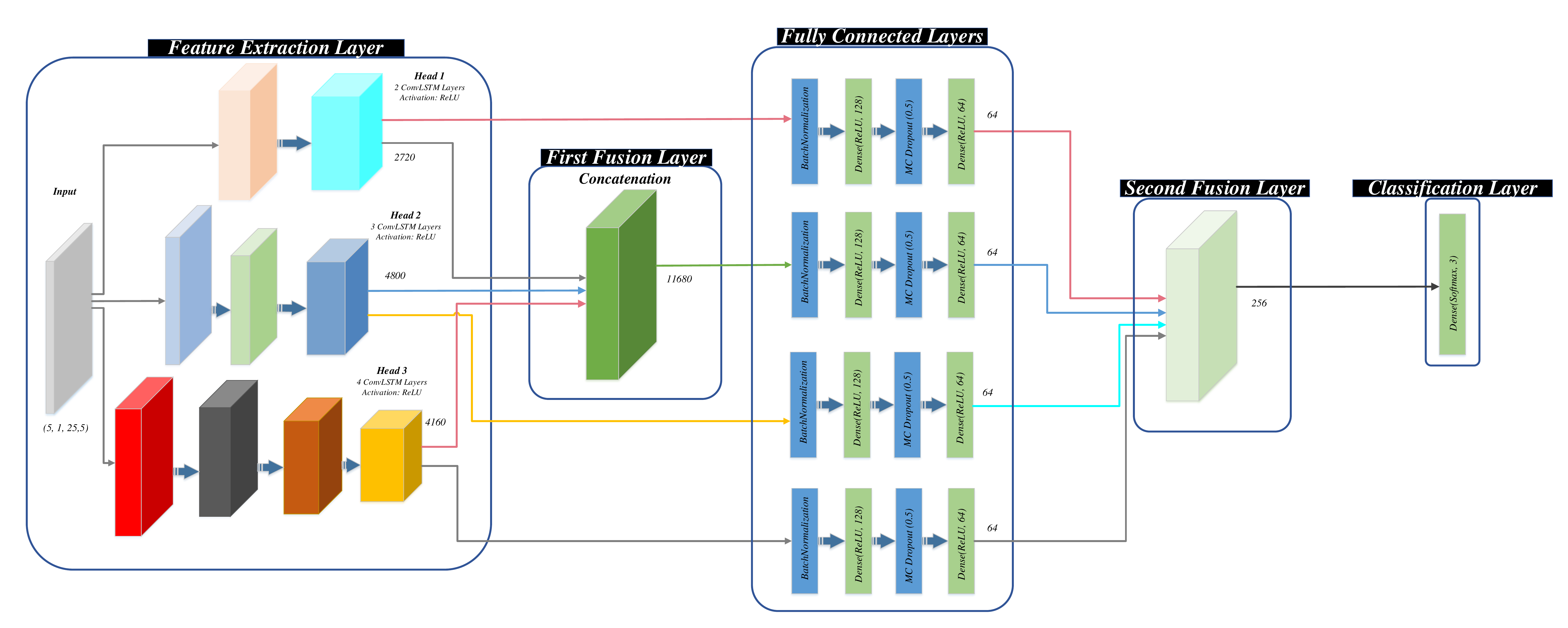}
     %\vspace*{0.1cm}
    \caption{A general overview of the proposed \emph{CASU2Net} model inspired by a novel feature fusion and dropout.}
    \label{CASU2Net_model}
\end{figure*}

The proposed \emph{CASU2Net} for fault detection is shown in Fig. \ref{CASU2Net_model}. In this model, We use ConvLSTM layers as feature extractors to benefit from its power in working with sequential data and convolution simultaneously. Each sample is a $125 \times 5$ matrix, and because of the input dimension in the input layer of ConvLSTM, each input has been reshaped to a 4-D matrix with $5\times 1\times25 \times 5$ dimensions. \ref{CASU2Net_model} consists of three parts. Two ConvLSTM layers form the first part with 32 and 32 filters and kernel sizes of $1 \times 5$ and $1 \times 5$, respectively. Three ConvLSTM layers form the second part with 32, 32, and 64 filters and kernel sizes of $1 \times 5$, $1 \times 5$, and $1 \times 5$, respectively. Finally, the third part is formed by four ConvLSTM layers with 32, 32, 64, and 64 filters with kernel sizes of $1 \times 5$, $1 \times 5$, $1 \times 3$, and $1 \times 3$, respectively. All ConvLSTM layers have ReLU as activation functions. In fact, each ConvLSTM layer has a precise meaning. For instance, when we use a filter with a size of $1 \times 5$, every 5 samples, the connection with other sensor samples will be calculated. The first fusion layer concatenates the outputs of each part and creates a feature pool from different sources that contains both high-level and simple features. After that, we have four fully connected layers part like the classification layers of simple CNN and multi-headed model, which get the output of the three parts and also the output of the first fusion layer. Each fully connected layers have a batch normalization layer, a dense layer with 128 hidden unit and ReLU as activation function, MC dropout with a rate equal to 0.5, and a dense layer with 64 hidden unit and ReLU as activation function. The second fusion layer merges four fully connected layers output, which creates a feature pool from processed data from the first fusion layer and three feature extraction parts. Indeed, the second fusion layer gives a comprehensive representation of the dataset to model and provides the model with advanced features to classify the signals. We finally have our output node connected to the dense layers with eight hidden units and softmax activation function for the classification algorithm.

\section{Experimental Results and Discussion}
In this section, the simulation results and discussion are presented. The proposed deep learning models for fault detection models are implemented using Keras library with Python 3.6 support \cite{ketkar2017introduction} and NVIDIA Tesla K80 model GPU. We also used Random Forest (RF) (max-depth=50,~n-estimators=200) and Decision Tree (DT) (max-depth=50) to evaluate their performance for fault detection.  After training the fault detection models, we use some metrics to evaluate and compare the performance of the proposed models with each other. The selected metrics are as follows \cite{abdar2019ne, zomorodi2021hybrid, abdar2019cwv}:

\begin{enumerate}

\item Accuracy=$\frac{TP+TN}{TP+TN+FP+FN}$

\item Precision=$\frac{TP}{TP+FP}$

\item Recall=$\frac{TP}{TP+FN}$

\item F-score=$\frac{2*TP}{2TP+FP+FN}$

\item Receiver operating characteristics (ROC)

\item Area under the ROC curve (AUC)

\item Confusion matrix,

\end{enumerate}
where TP stands for true positive (correctly classified), TN stands for true negative (correctly classified), FP stands for false positive (incorrectly classified) and FN stands for false negative (incorrectly classified). More information about accuracy, precision, recall, F-score, ROC, AUC and confusion matrix can be found in \cite{bishop2006pattern}, \cite{fawcett2004roc}, \cite{he2009learning}. In this paper, we will obtain all of the metrics with 10-fold cross-validation.

\subsection{Fault detection without considering uncertainty}
\label{FDwithout}

\begin{figure*}[h!]
\centering
    \begin{subfigure}[b]{0.32\textwidth}
            \includegraphics[width=\textwidth]{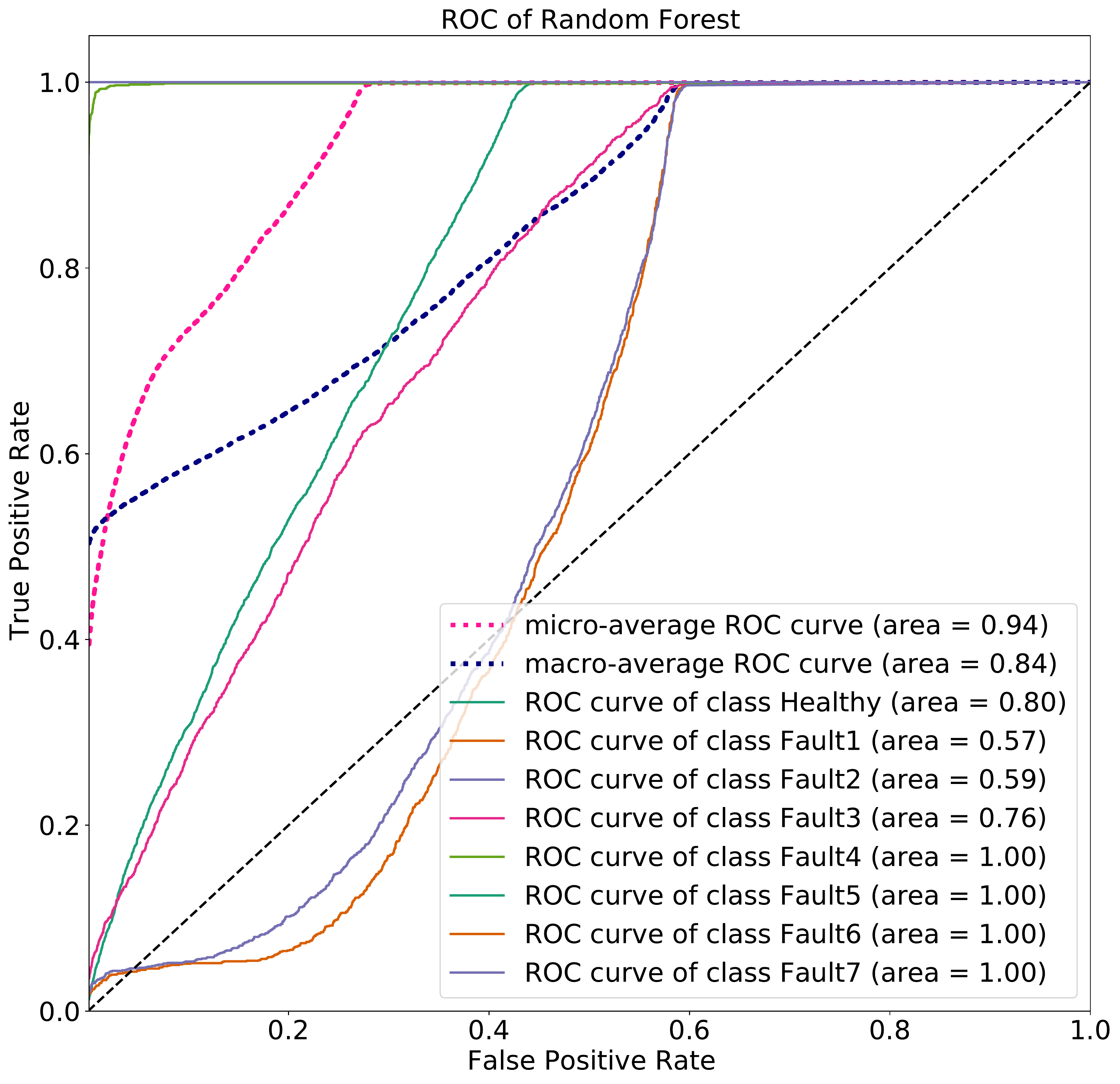}
            \caption{Random Forest}
            \label{ROC:RF}
    \end{subfigure}
       %\vspace*{0.5cm}
    \begin{subfigure}[b]{0.32\textwidth}
            \centering
            \includegraphics[width=\textwidth]{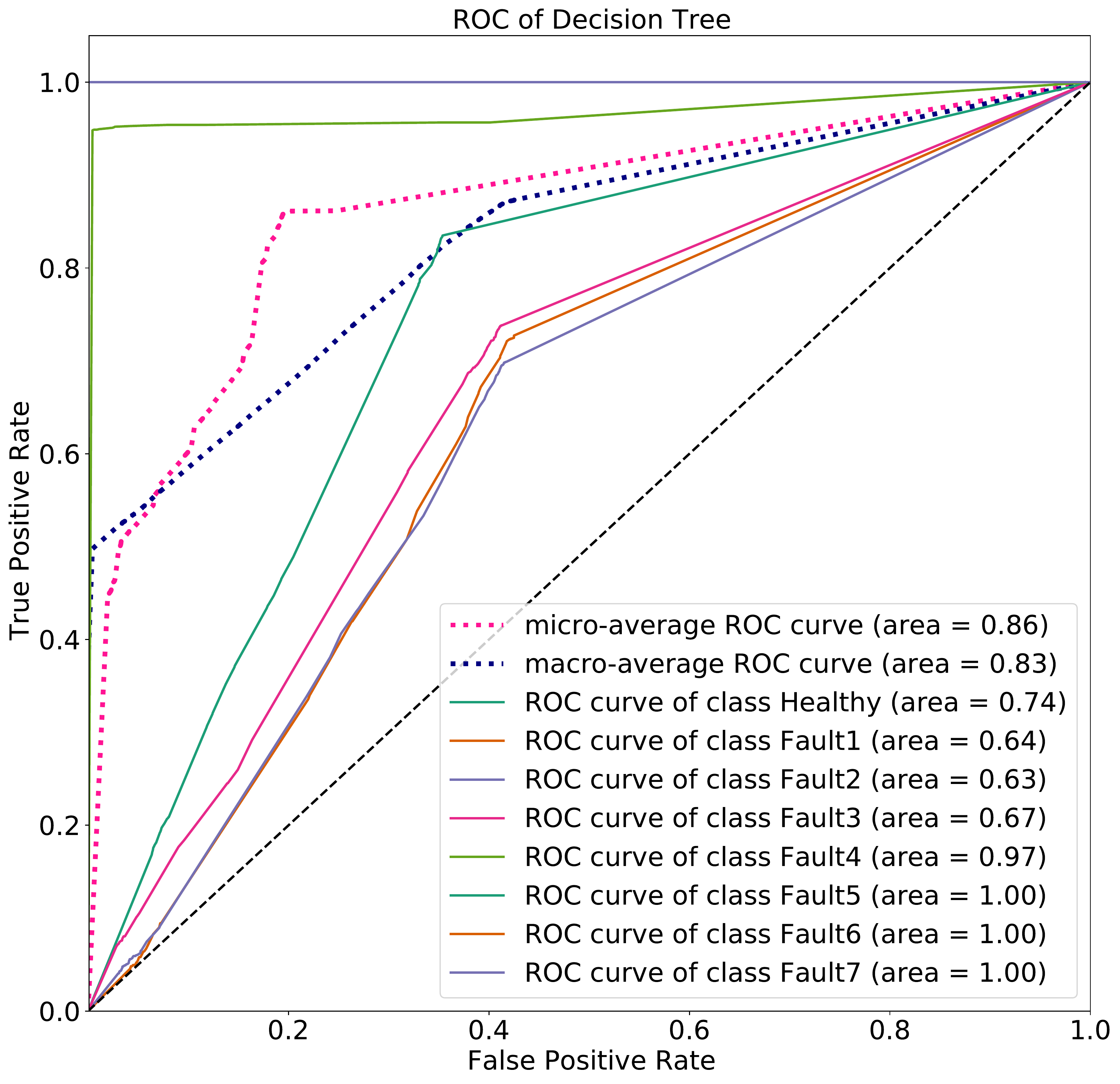}
            \caption{Decision Tree}
            \label{ROC:DT}
    \end{subfigure} 
    %\vspace*{0.5cm}
        \begin{subfigure}[b]{0.32\textwidth}
            \centering
            \includegraphics[width=\textwidth]{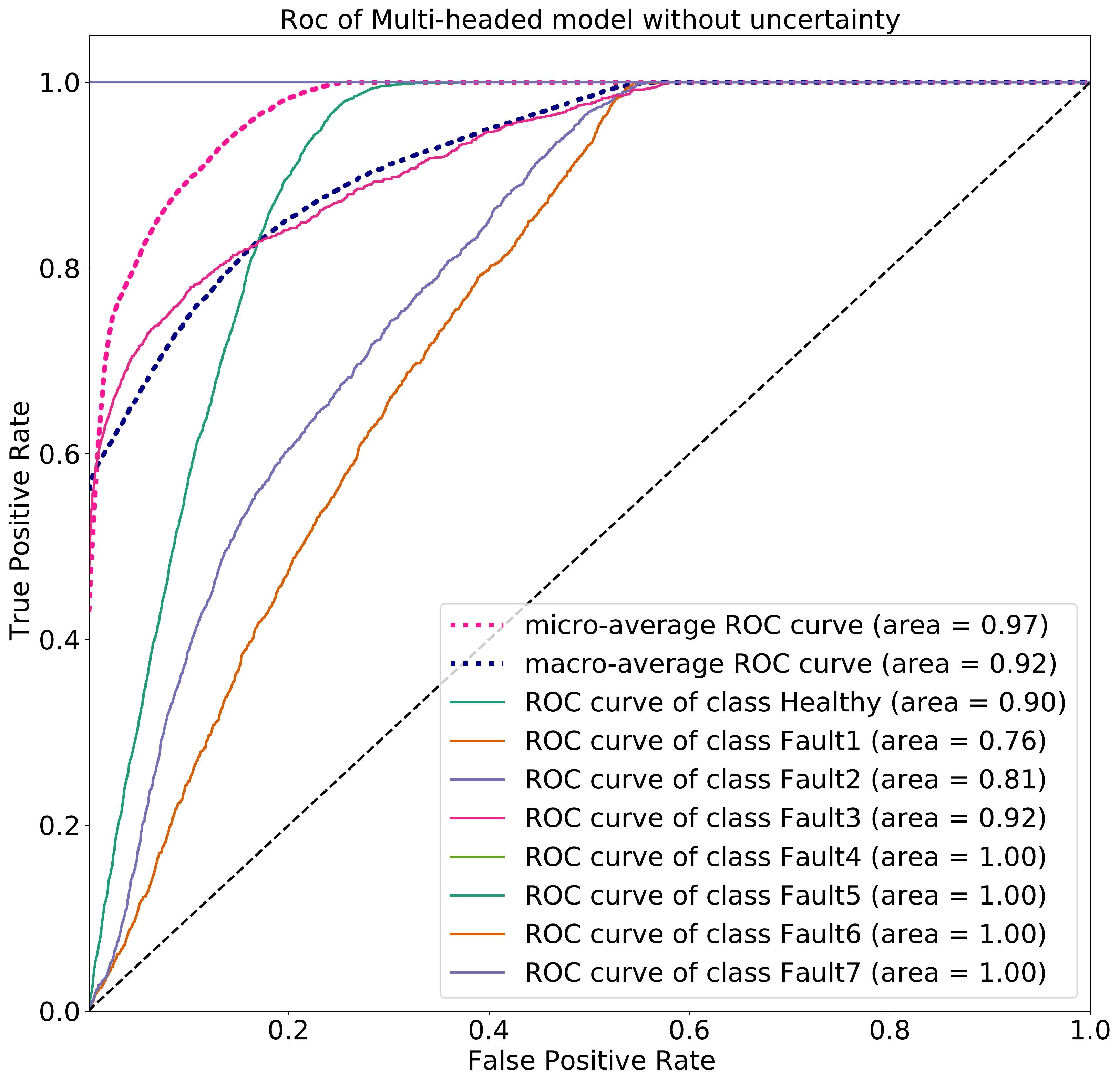}
            \caption{Multi-headed model without uncertainty}
            \label{ROC:MH_without}
    \end{subfigure} \\
        \begin{subfigure}[b]{0.32\textwidth}
            \centering
            \includegraphics[width=\textwidth]{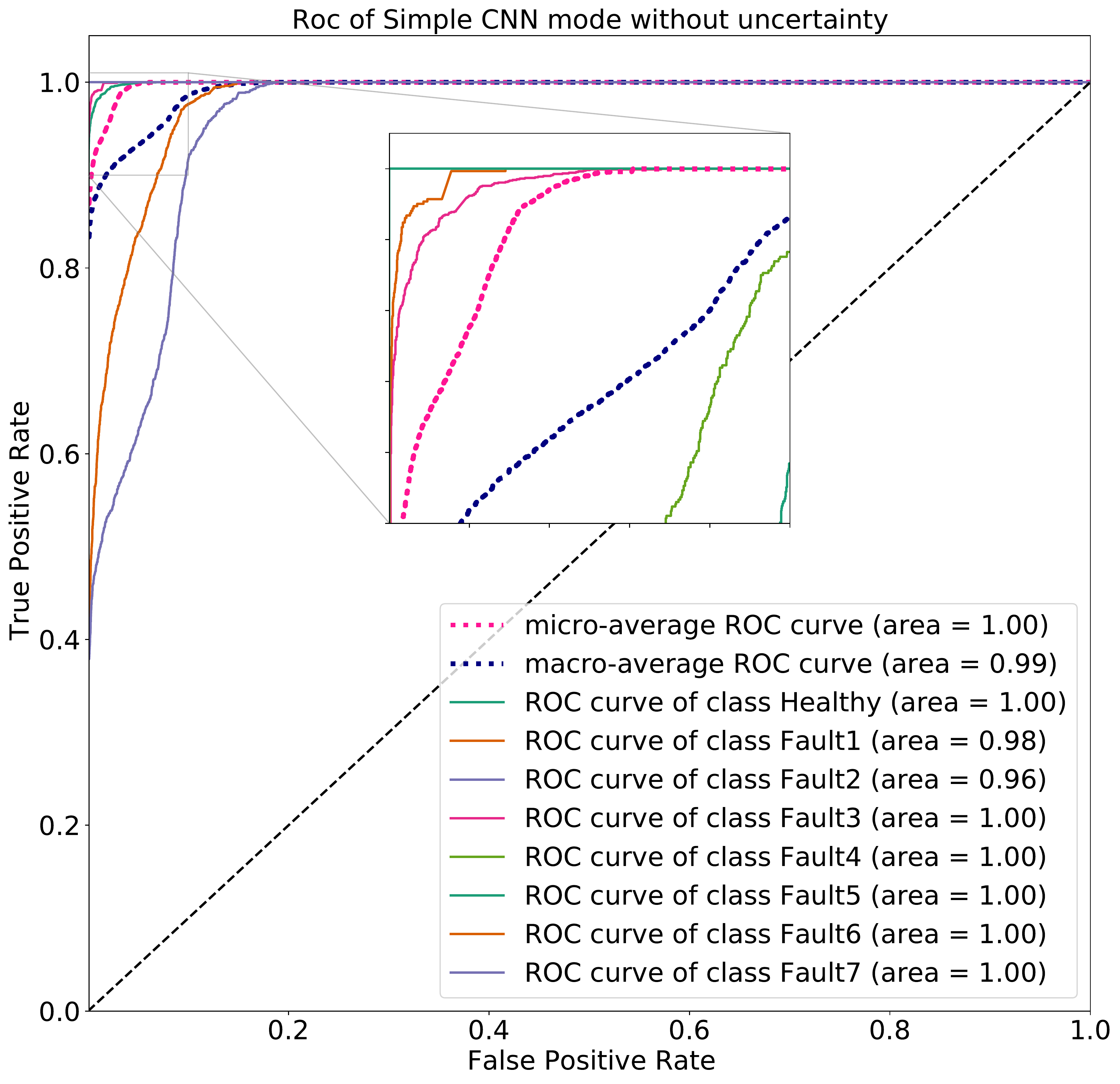} 
            \caption{Simple CNN without Uncertainty}
            \label{ROC:Simp_without}
    \end{subfigure} 
    \begin{subfigure}[b]{0.32\textwidth}
            \centering
            \includegraphics[width=\textwidth]{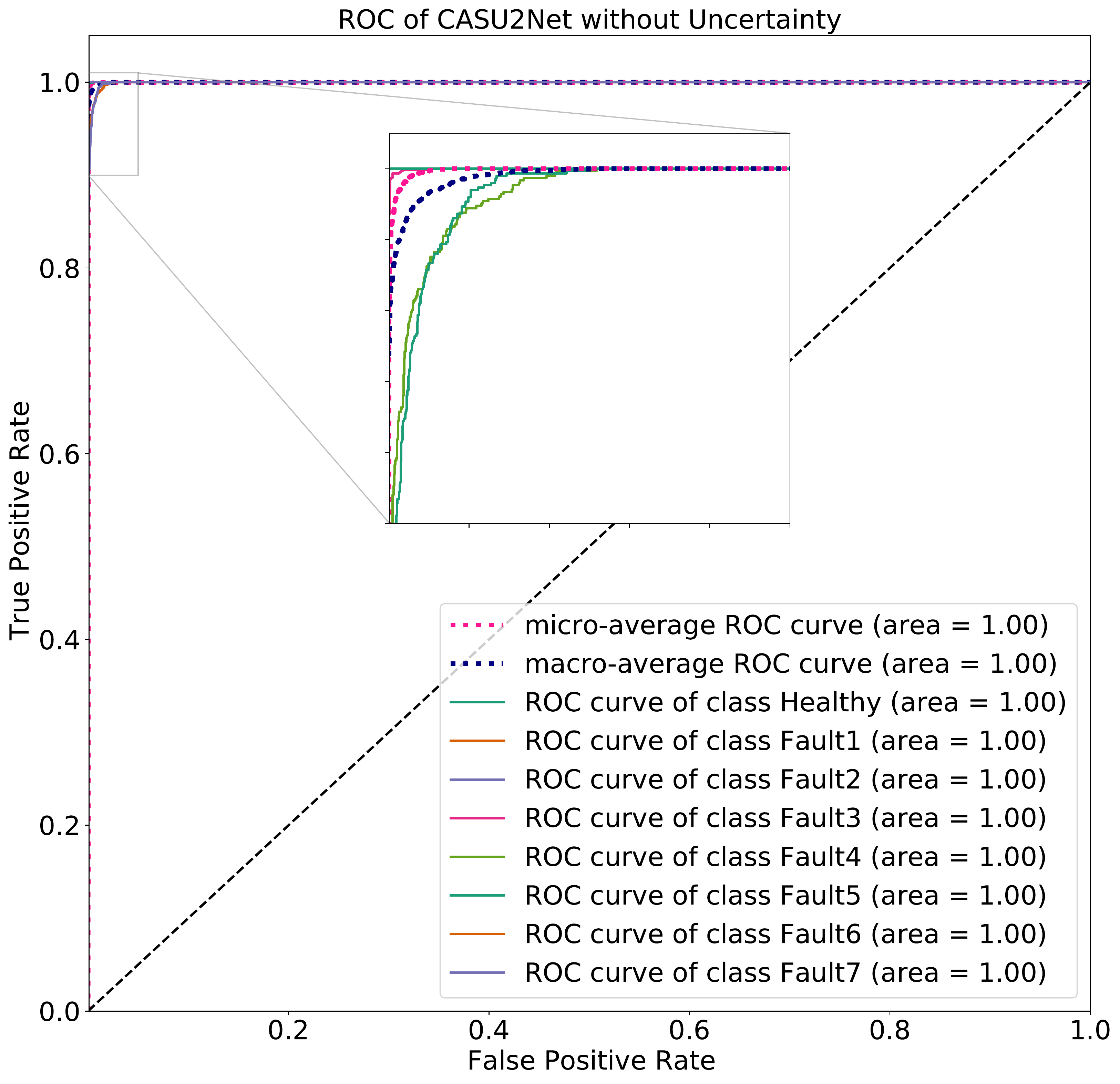}
            \caption{\emph{CASU2Net} without Uncertainty}
            \label{CASU2Net_without}
    \end{subfigure}
    \caption{ROC obtained without uncertainty quantification.}\label{ROC_Without}

\end{figure*}

\begin{figure*}[hbt!] 
\centering
    \begin{subfigure}[b]{0.2\textwidth}
            \includegraphics[width=\textwidth]{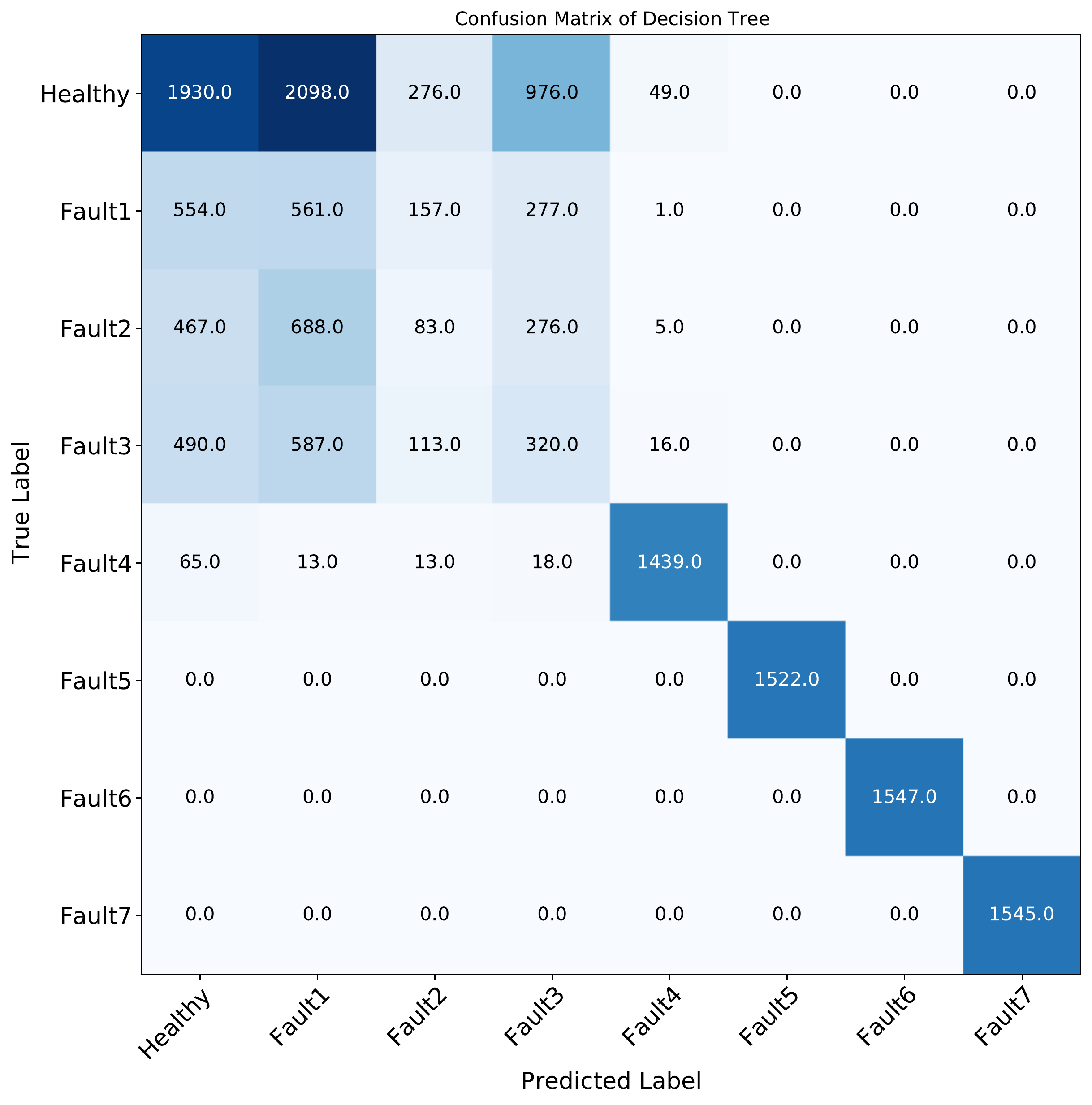}
            \caption{Decision Tree}
            \label{fig:SSL_F11111}
    \end{subfigure}
       %\vspace*{0.5cm}
    \begin{subfigure}[b]{0.2\textwidth}
            \centering
            \includegraphics[width=\textwidth]{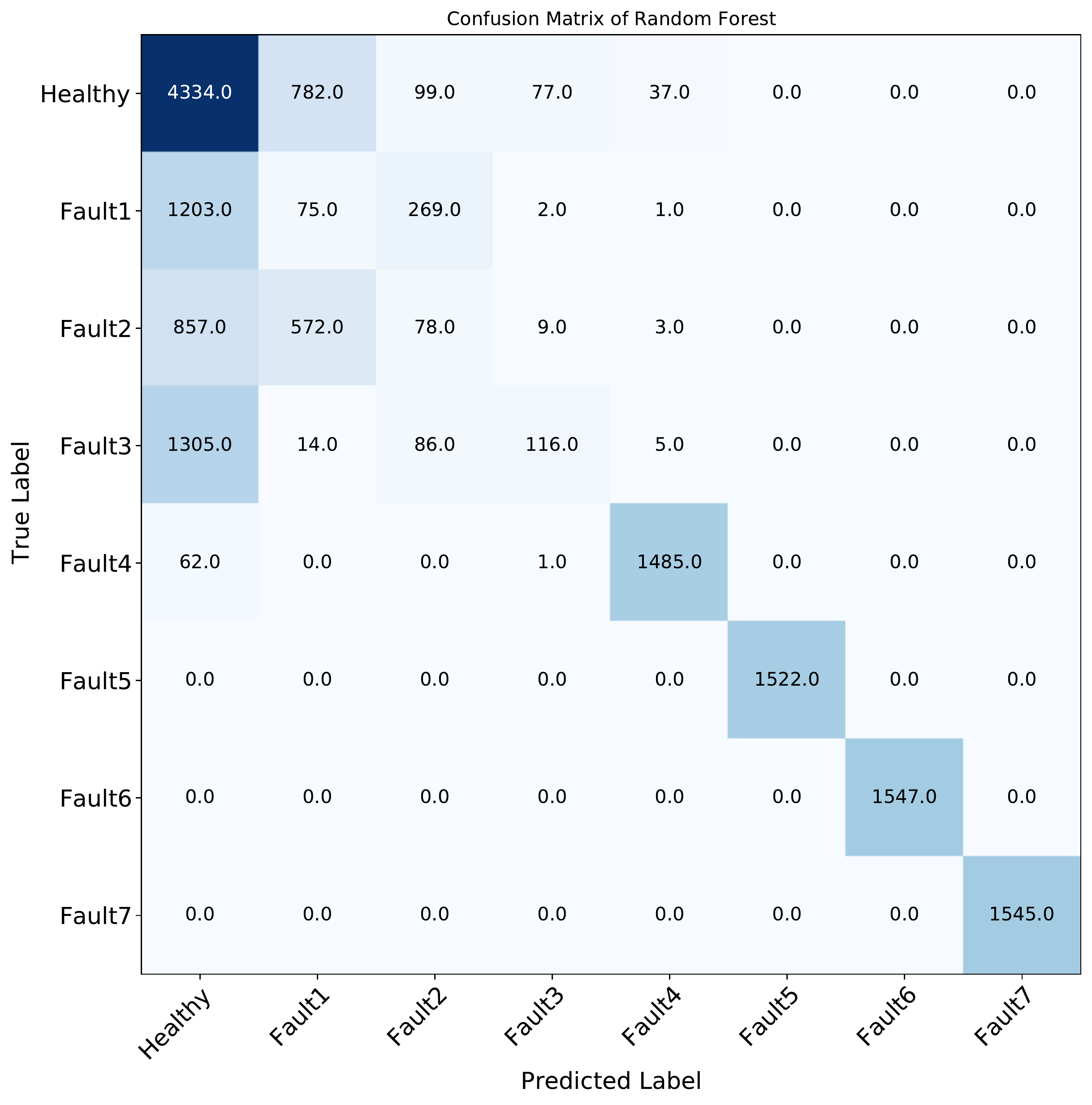}
            \caption{Random Forest}
            \label{fig:SSL_F222}
    \end{subfigure} 
    %\vspace*{0.5cm}
    \begin{subfigure}[b]{0.2\textwidth}
            \centering
            \includegraphics[width=\textwidth]{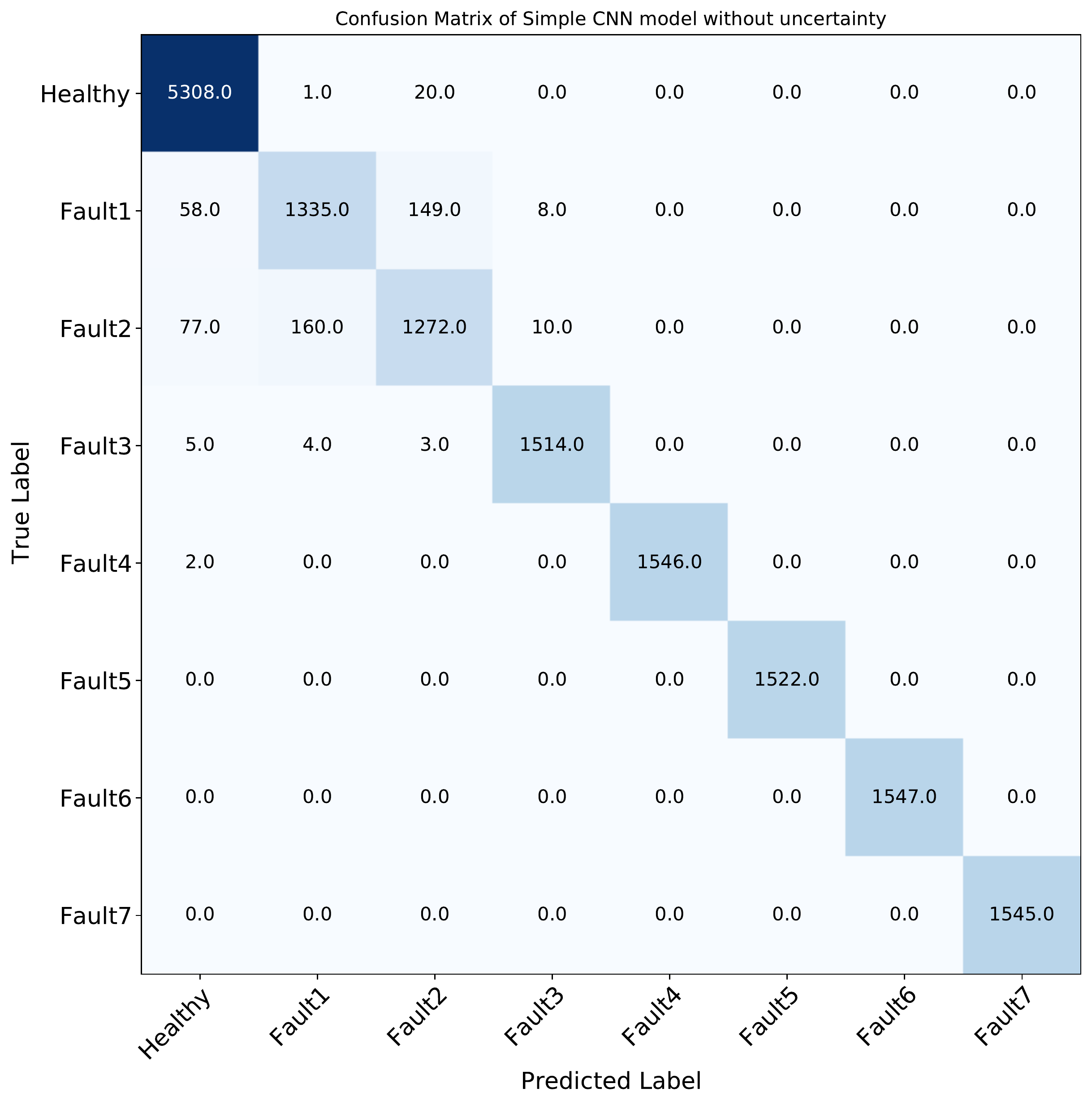}
            \caption{Simple CNN}
            \label{fig:SSL_F24}
    \end{subfigure}
    \begin{subfigure}[b]{0.2\textwidth}
            \centering
            \includegraphics[width=\textwidth]{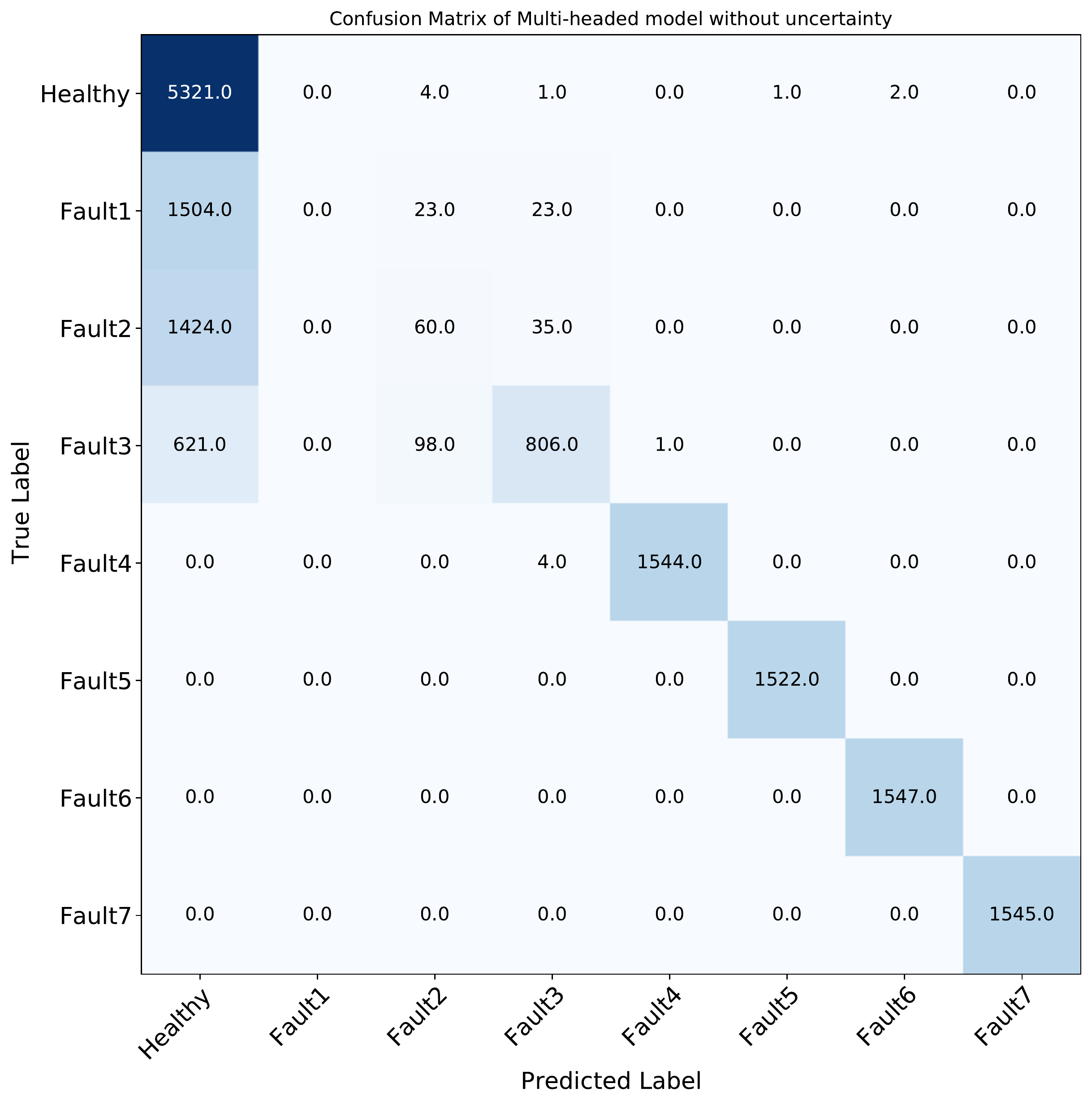}
            \caption{Multi-headed}
            \label{fig:1}
    \end{subfigure}  
    \begin{subfigure}[b]{0.2\textwidth}
            \centering
            \includegraphics[width=\textwidth]{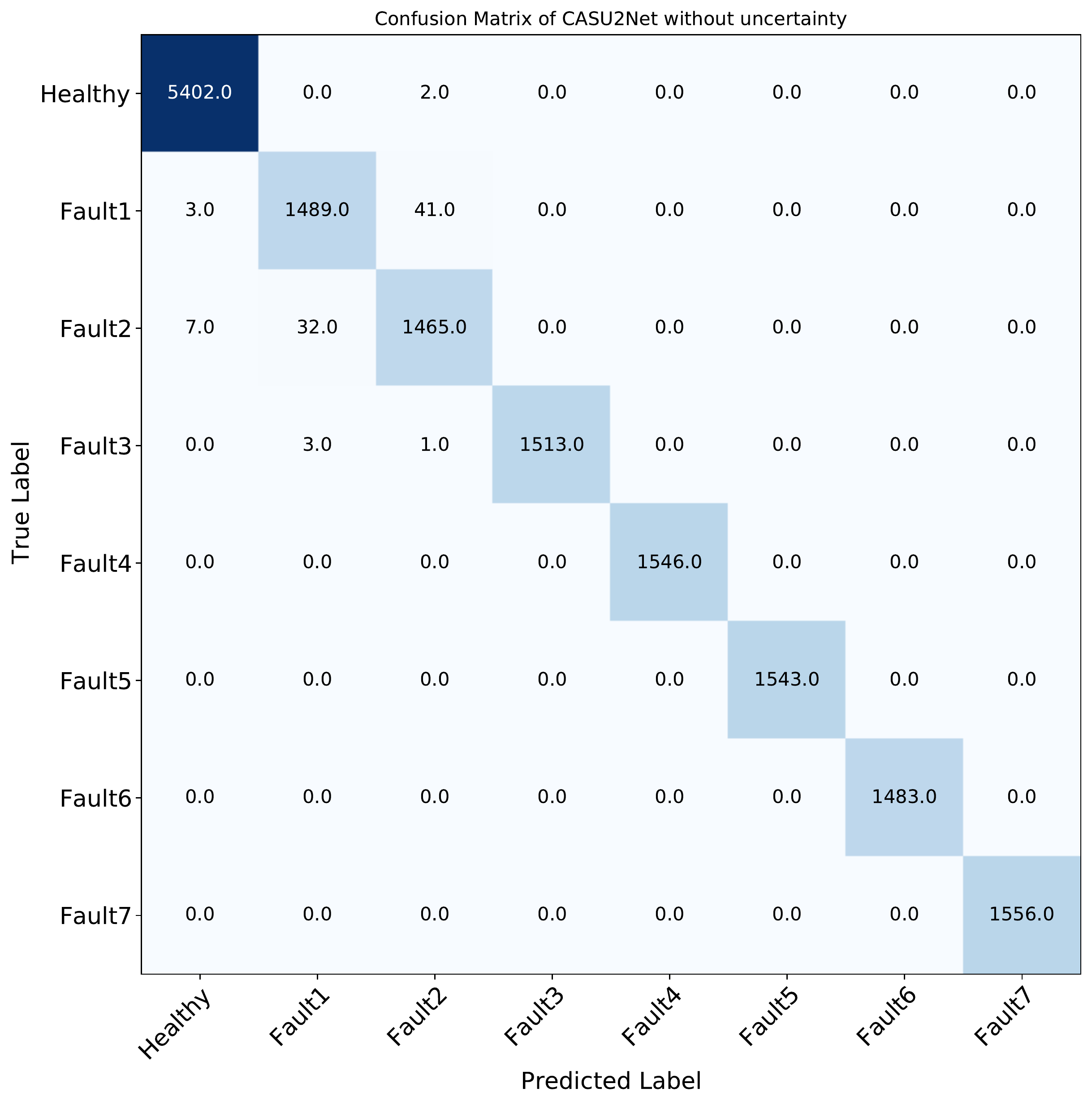}
            \caption{\emph{CASU2Net}}
            \label{fig:2}
    \end{subfigure}
     %\vspace*{0.1cm}
         \caption{Confusion matrices obtained without uncertainty quantification.}\label{CMWithout}
\end{figure*}

First, we examined the results of fault detection models including RF, DT, simple CNN, multi-headed, and our proposed fusion model (\emph{CASU2Net}) without considering uncertainty. The acquired results are presented in Table \ref{Result_WITOUTUQ}. We can see from Table \ref{Result_WITOUTUQ} that proposed \emph{CASU2Net} outperformed the other implemented fault detection models with a test accuracy of 99.4\%. The results also indicate that DT has reached the weakest performance among all of the utilized models. The ROC curves and confusion matrices for models without uncertainty quantification are also shown in Figs. \ref{ROC_Without} and \ref{CMWithout}.

\subsection{Fault detection with considering uncertainty}

The achieved outcomes of the fault detection models from sub-section (\ref{FDwithout}) are prominent; hence, our \emph{CASU2Net} can be employed in real-world applications to reduce the offshore wind turbine operational cost and enhance their maintenance. But at the same time, we strongly believe in the transparency of the results obtained by such intelligent-based models with certainty. To accomplish this, we employed the uncertainty quantification technique named MC dropout to evaluate the uncertainty in simple CNN, multi-headed model, and \emph{CASU2Net}. 

\begin{figure*}[h!]
\centering
        \begin{subfigure}[b]{0.32\textwidth}
            \centering
            \includegraphics[width=\textwidth]{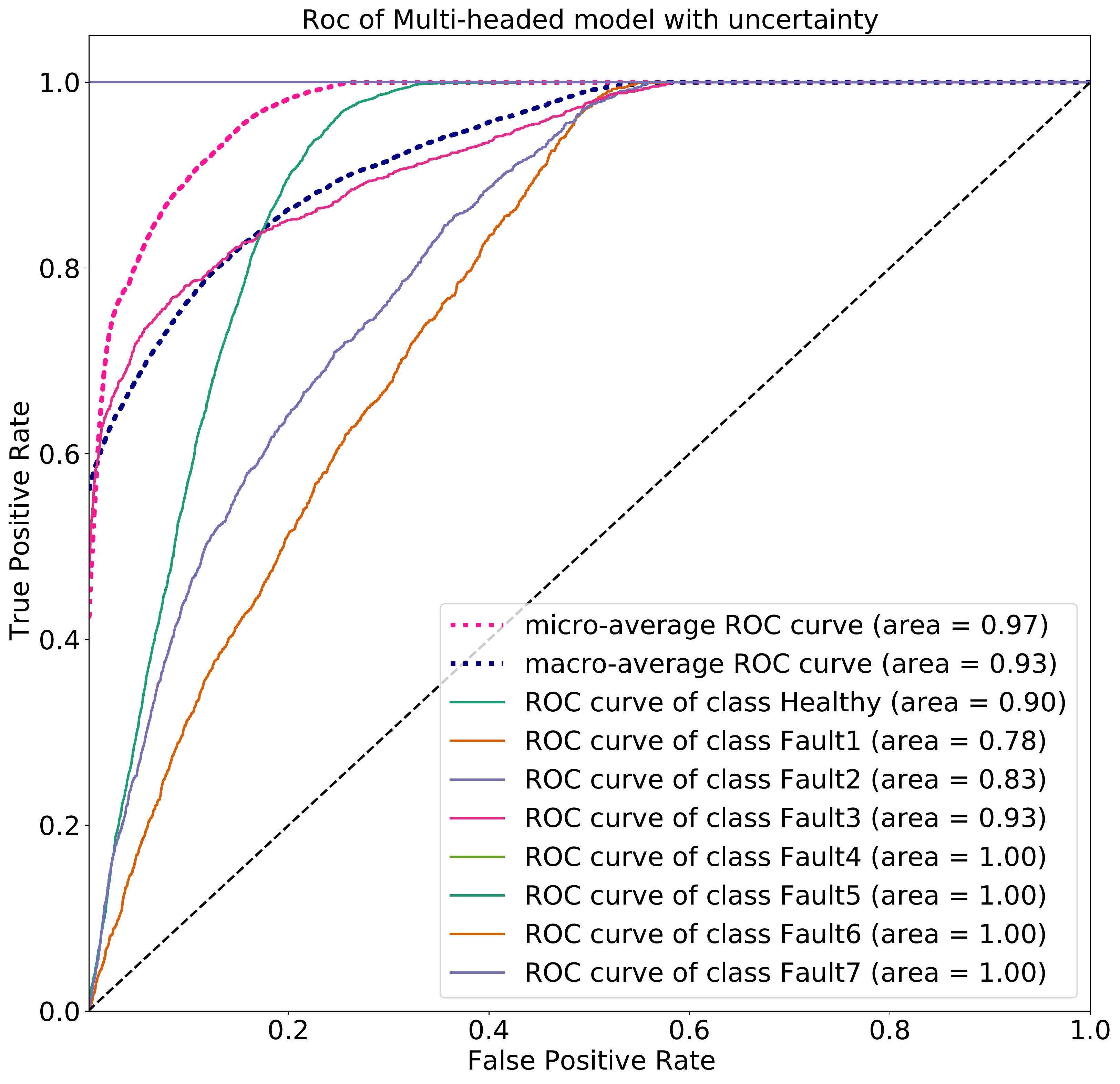}
            \caption{Multi-headed model with uncertainty}
            \label{ROC:MH_with}
    \end{subfigure}
        \begin{subfigure}[b]{0.32\textwidth}
            \centering
            \includegraphics[width=\textwidth]{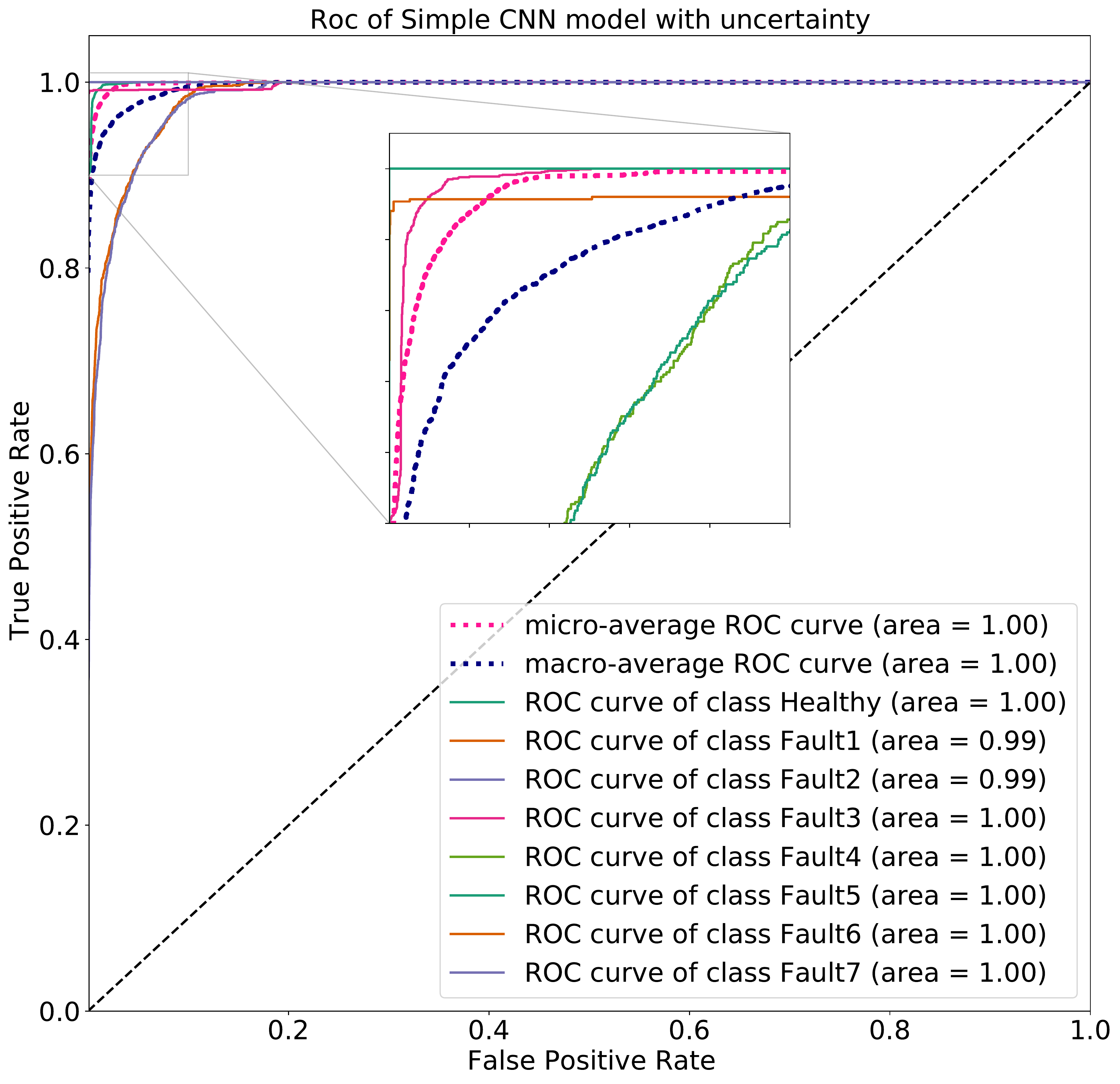}
            \caption{Simple CNN with Uncertainty}
            \label{ROC:Simp_with}
    \end{subfigure} 
    \begin{subfigure}[b]{0.32\textwidth}
            \centering
            \includegraphics[width=\textwidth]{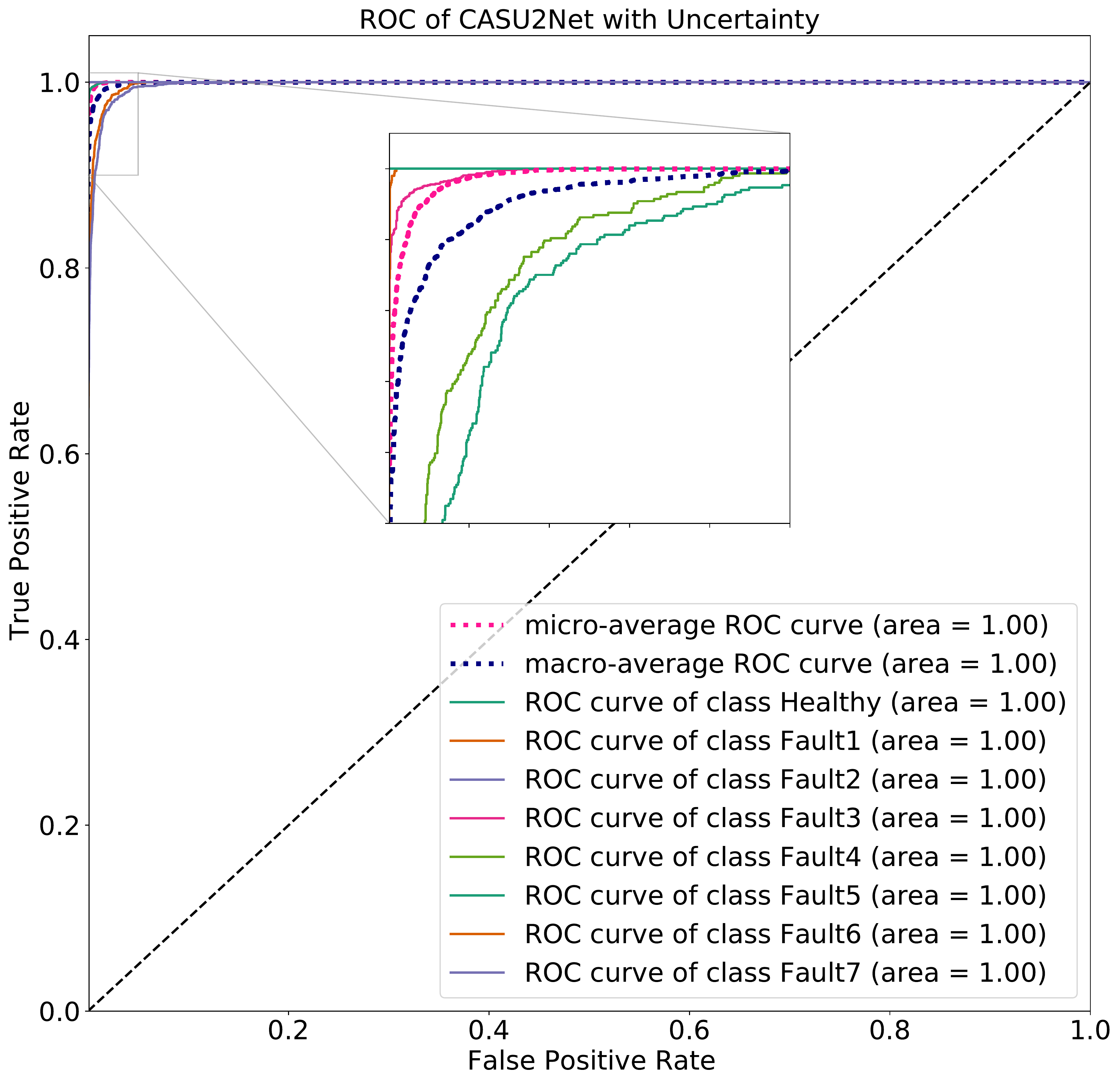}
            \caption{\emph{CASU2Net} with Uncertainty}
            \label{CASU2Net_with}
    \end{subfigure}\\ 
    \caption{ROC obtained with uncertainty quantification.}\label{ROC_With}
\end{figure*}

\begin{figure*}[!ht]
\centering
    \begin{subfigure}[b]{0.2\textwidth}
            \centering
            \includegraphics[width=\textwidth]{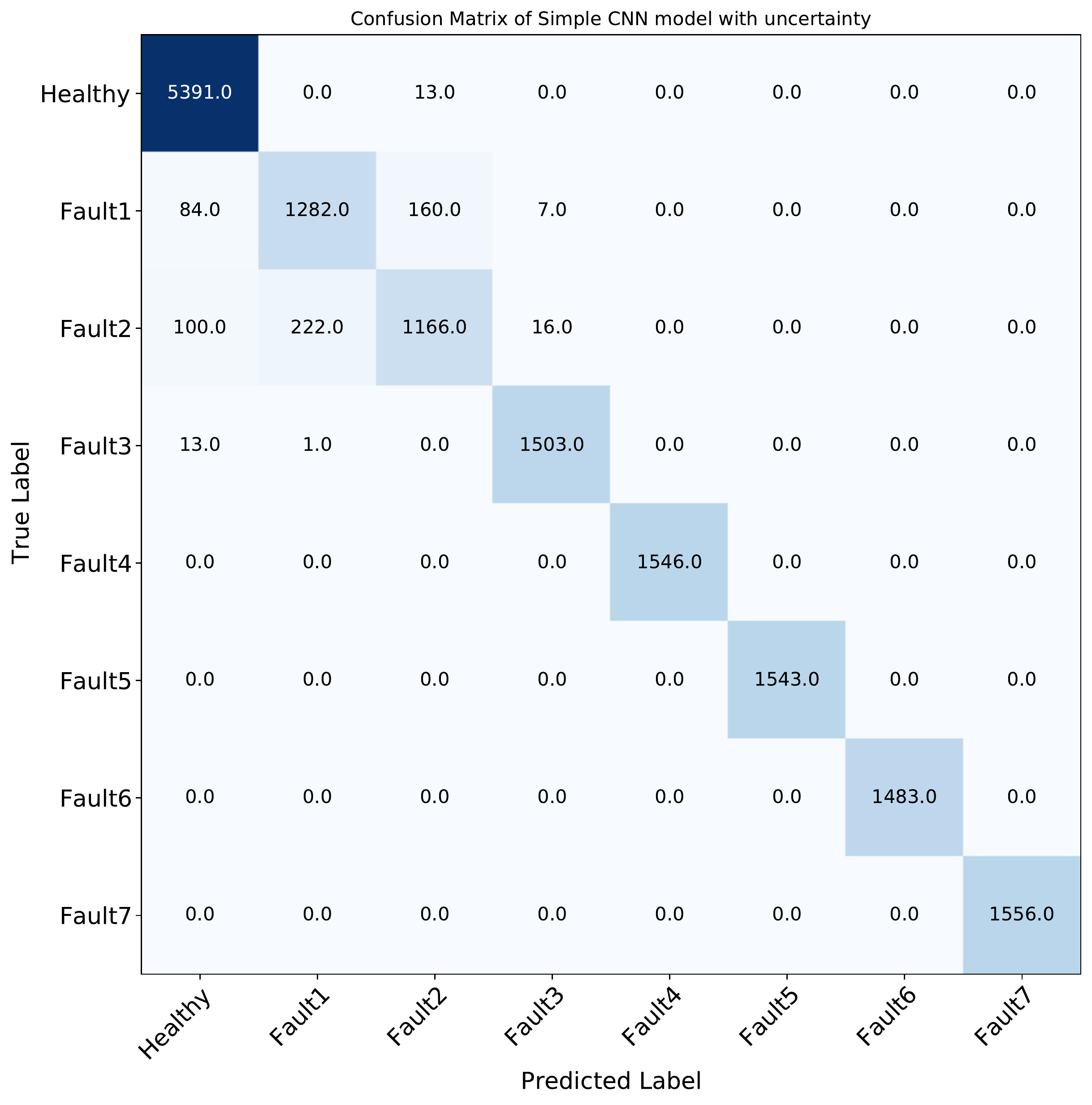}
            \caption{Simple CNN}
            \label{fig:SSL_F22}
    \end{subfigure}
    \begin{subfigure}[b]{0.2\textwidth}
            \centering
            \includegraphics[width=\textwidth]{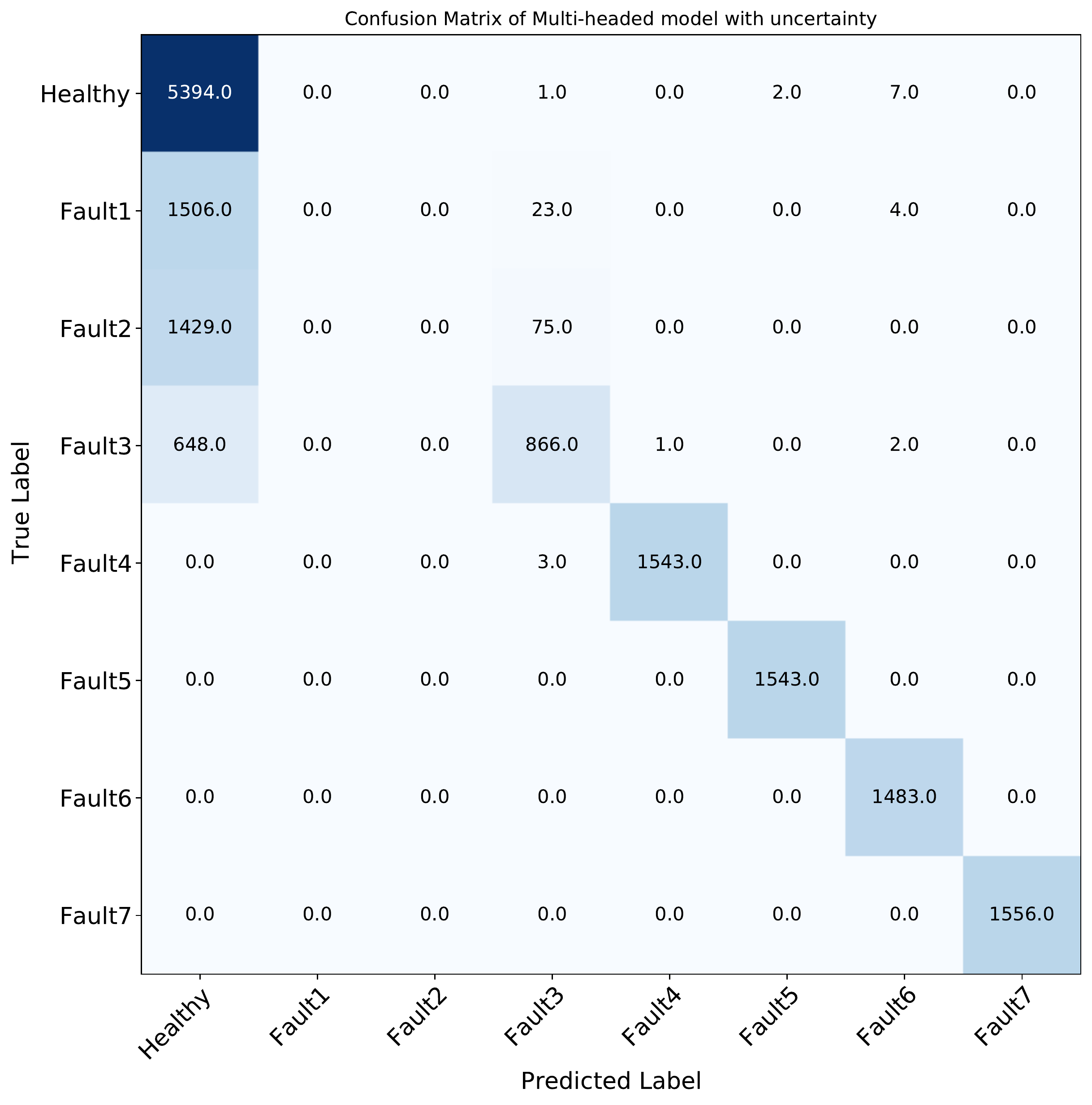}
            \caption{Multi-headed}
            \label{fig:5}
    \end{subfigure}  
    \begin{subfigure}[b]{0.2\textwidth}
            \centering
            \includegraphics[width=\textwidth]{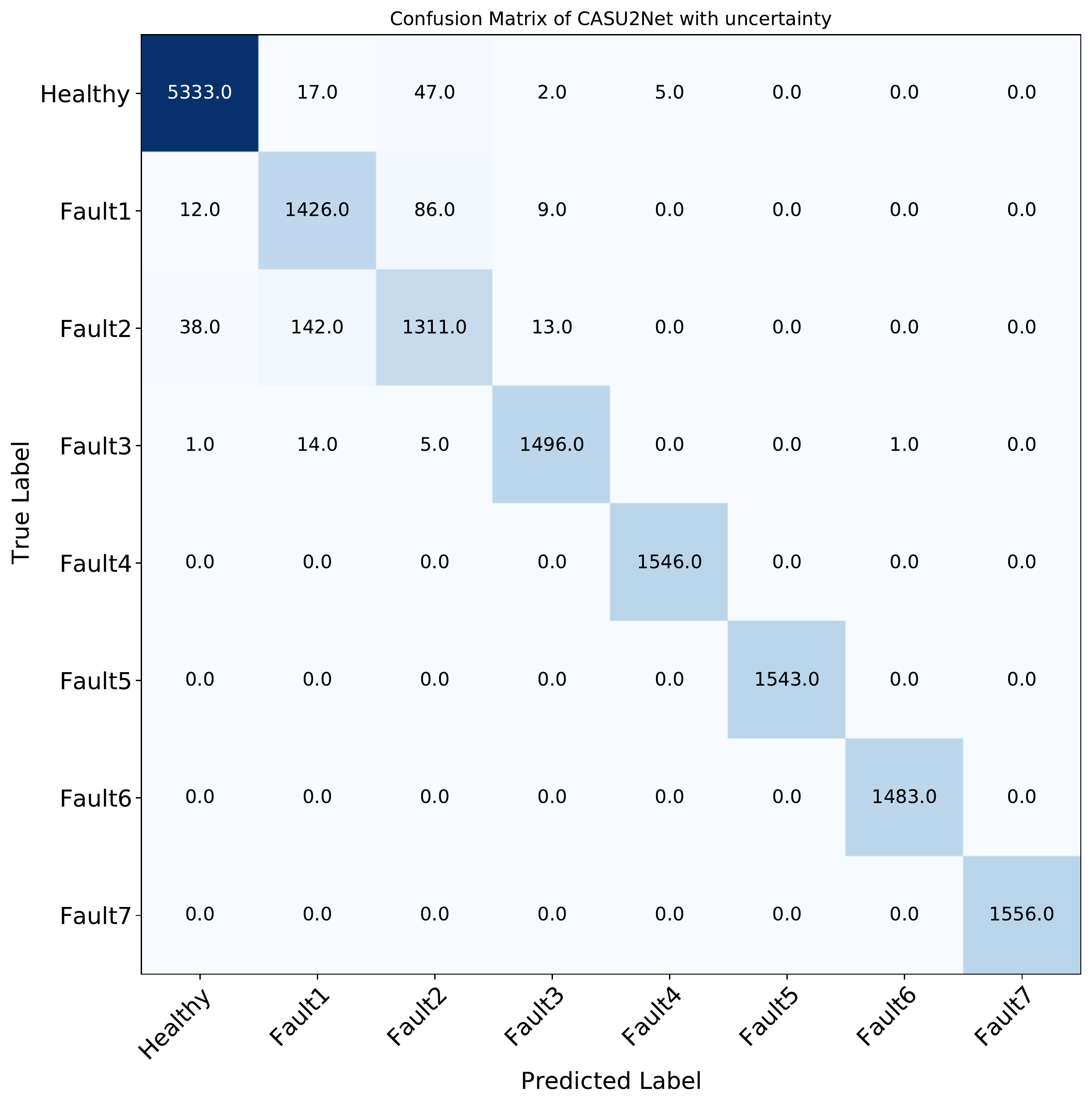}
            \caption{\emph{CASU2Net}}
            \label{fig:66}
    \end{subfigure}
     %\vspace*{0.1cm}
    \caption{Confusion matrices obtained with uncertainty quantification.}\label{CMWith}
\end{figure*}

Table \ref{Results_WithUQ} and Figs. \ref{ROC_With}, and \ref{CMWith} demonstrate results, ROC curves, and confusion matrices of the fault detection models with uncertainty quantification. We can perceive from Table \ref{Results_WithUQ} that our \emph{CASU2Net} obtained the best results with an accuracy of 98.2\%. Comparing the results obtained from deep learning models without and with the UQ method, we found that our \emph{CASU2Net} and multi-headed model with the UQ method have achieved imperceptibly weaker performance than the models without UQ. But, simple CNN with UQ worked better than the model without UQ.

%Table \ref{tab_result} presents the results of 10-fold cross-validation. The confusion matrices for models are also shown in Figs. \ref{CMWithout} and \ref{CMWith}. Moreover, the ROC curves for models are shown in Figs. \ref{ROC_Without} and \ref{ROC_With}. %In ROC curves, class 0 is for normal samples, while other classes (classes 1-7) are for fault samples. 
%When working with time series data, because the patterns change over time, the models should demonstrate a good performance at all times. The network architecture and parameters of the models are tuned by trial and error, based on practical knowledge and experience.

\begin{table}[h]
%\hspace{-30cm}
\caption{Results of each model without considering uncertainty.}
\centering
\begin{tabular}{l l l l l l}
\hline
Model & Accuracy & Precision & Recall & F-score \\
\hline
\hline
Decision Tree & 54.1\% & 0.607 & 0.541 & 0.572 \\
Random Forest & 66.7\% & 0.632 & 0.667 & 0.649 \\
Deep 1 (Simple CNN) & 93.5\% & 0.934 & 0.935 & 0.937 \\
Deep 2 (Multi-headed) & 77.2\% & 0.708 & 0.772 & 0.738 \\
\emph{CASU2Net} & \textbf{99.4}\% & \textbf{0.994} &\textbf{0.994} & \textbf{0.994}\\

\hline
\end{tabular}
\label{Result_WITOUTUQ}
\end{table}

\begin{table}[h]
%\hspace{-30cm}
\caption{Results of each model with considering uncertainty.}
\centering
\begin{tabular}{l l l l l l}
\hline
Model & Accuracy & Precision & Recall & F-score \\
\hline
\hline

Deep 1 (Simple CNN) & 96.2\% & 0.96 & 0.962 & 0.96 \\
Deep 2 (Multi-headed) & 77\% & 0.666 & 0.77 & 0.714 \\
\emph{CASU2Net} & \textbf{98.2}\% & \textbf{0.983} &\textbf{0.982} & \textbf{0.983}\\

\hline
\end{tabular}
\label{Results_WithUQ}
\end{table}

\subsection{Layer representation}

\begin{figure*}[h!]
\centering
    \begin{subfigure}[b]{0.3\textwidth}
            \centering
            \includegraphics[width=\textwidth]{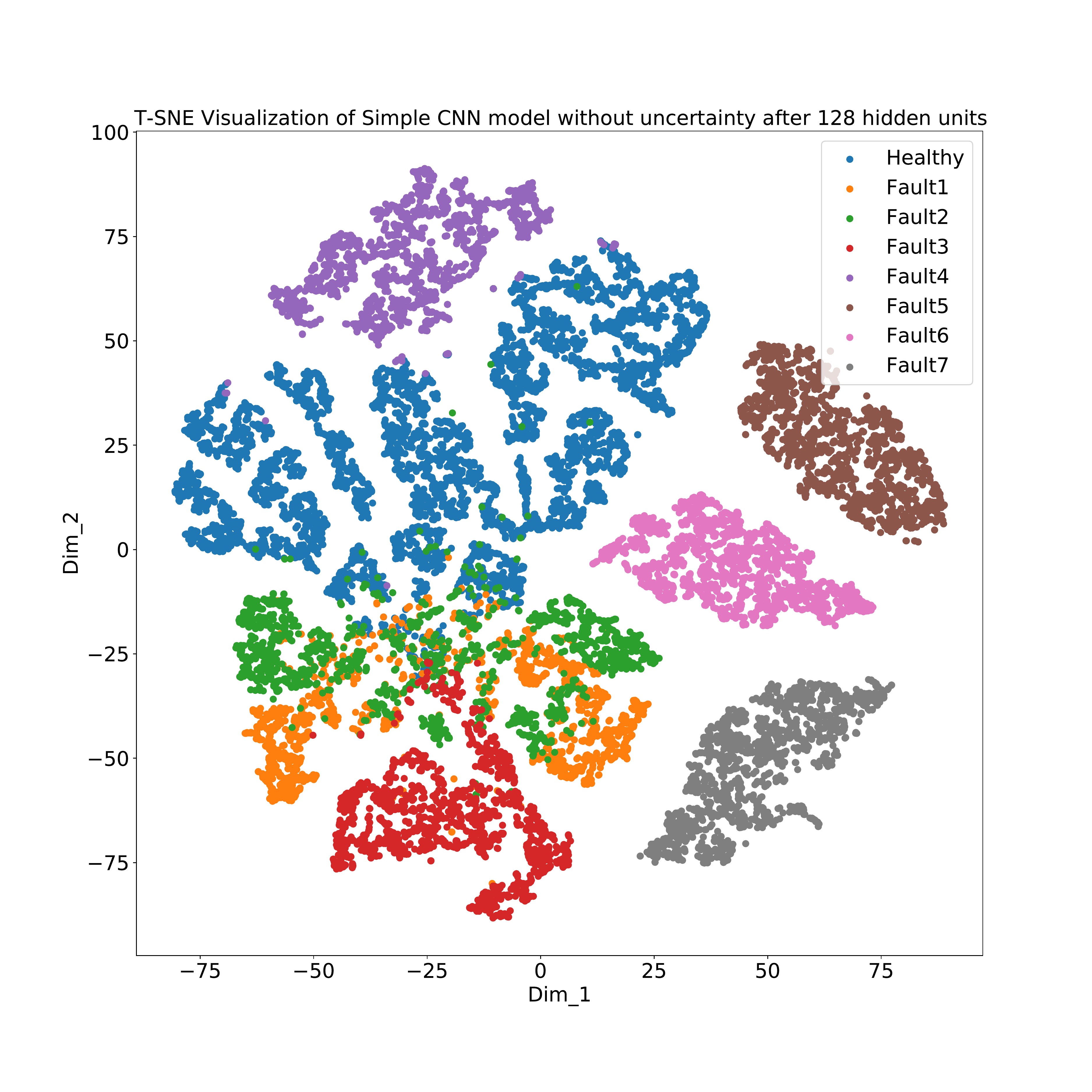}
            \caption{Simple CNN without UQ}
            \label{TSNE_Simple_without}
    \end{subfigure}
    \begin{subfigure}[b]{0.3\textwidth}
            \centering
            \includegraphics[width=\textwidth]{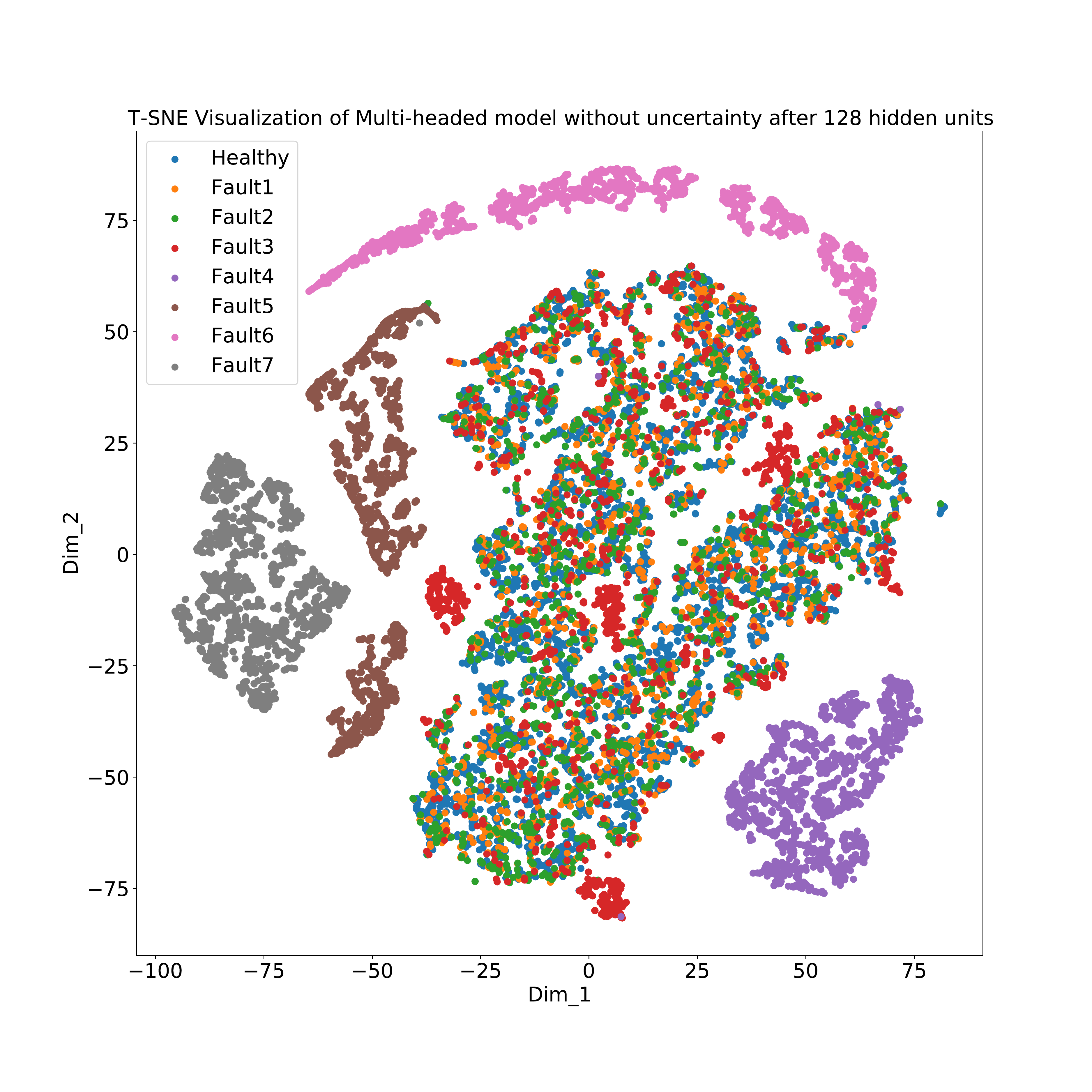}
            \caption{Multi-headed without UQ}
            \label{TSNE_Multi_without}
    \end{subfigure}  
    \begin{subfigure}[b]{0.3\textwidth}
            \centering
            \includegraphics[width=\textwidth]{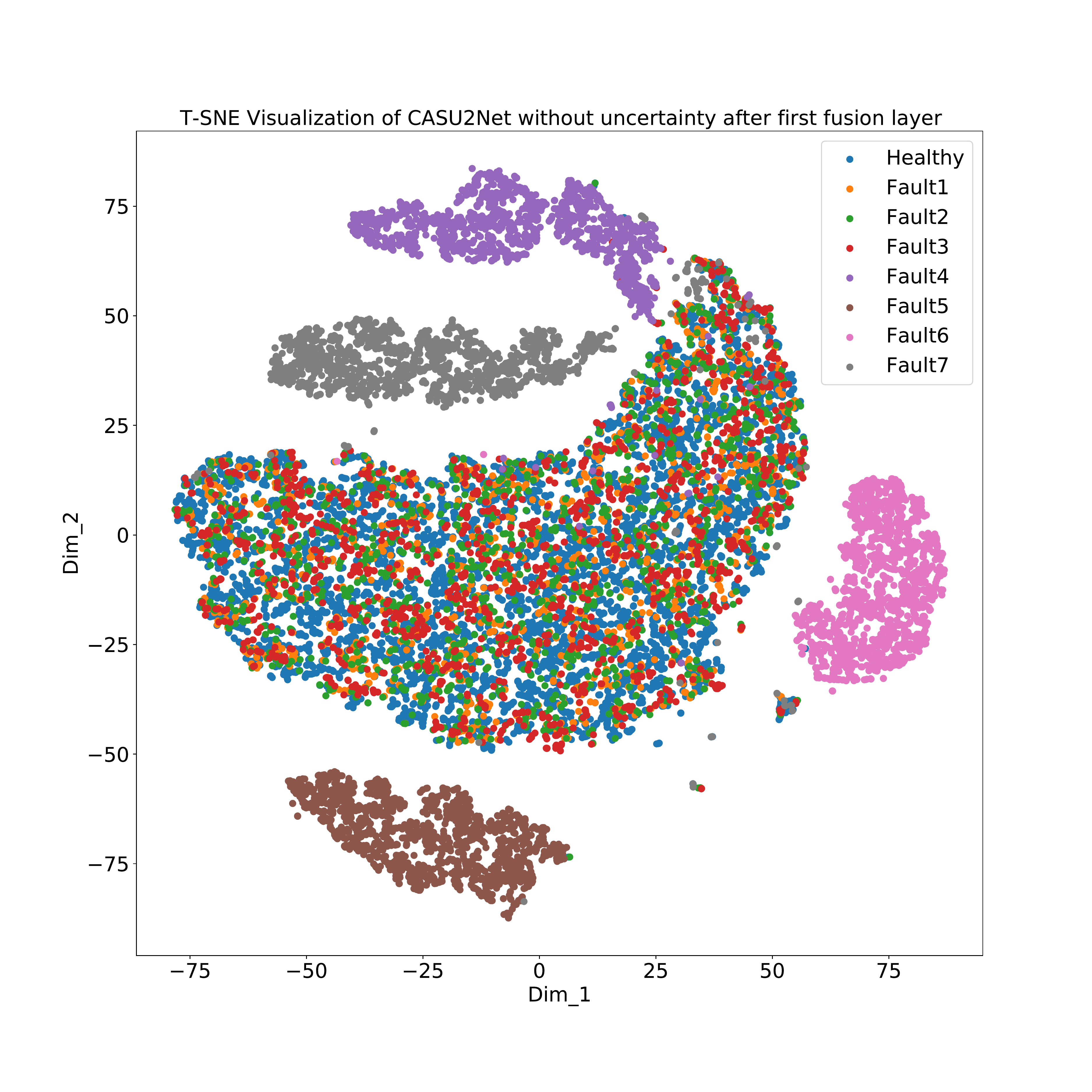}
            \caption{\emph{CASU2Net} first fusion layer without UQ}
            \label{TSNE_CASU2Net_without_first}
    \end{subfigure}
     \begin{subfigure}[b]{0.3\textwidth}
            \centering
            \includegraphics[width=\textwidth]{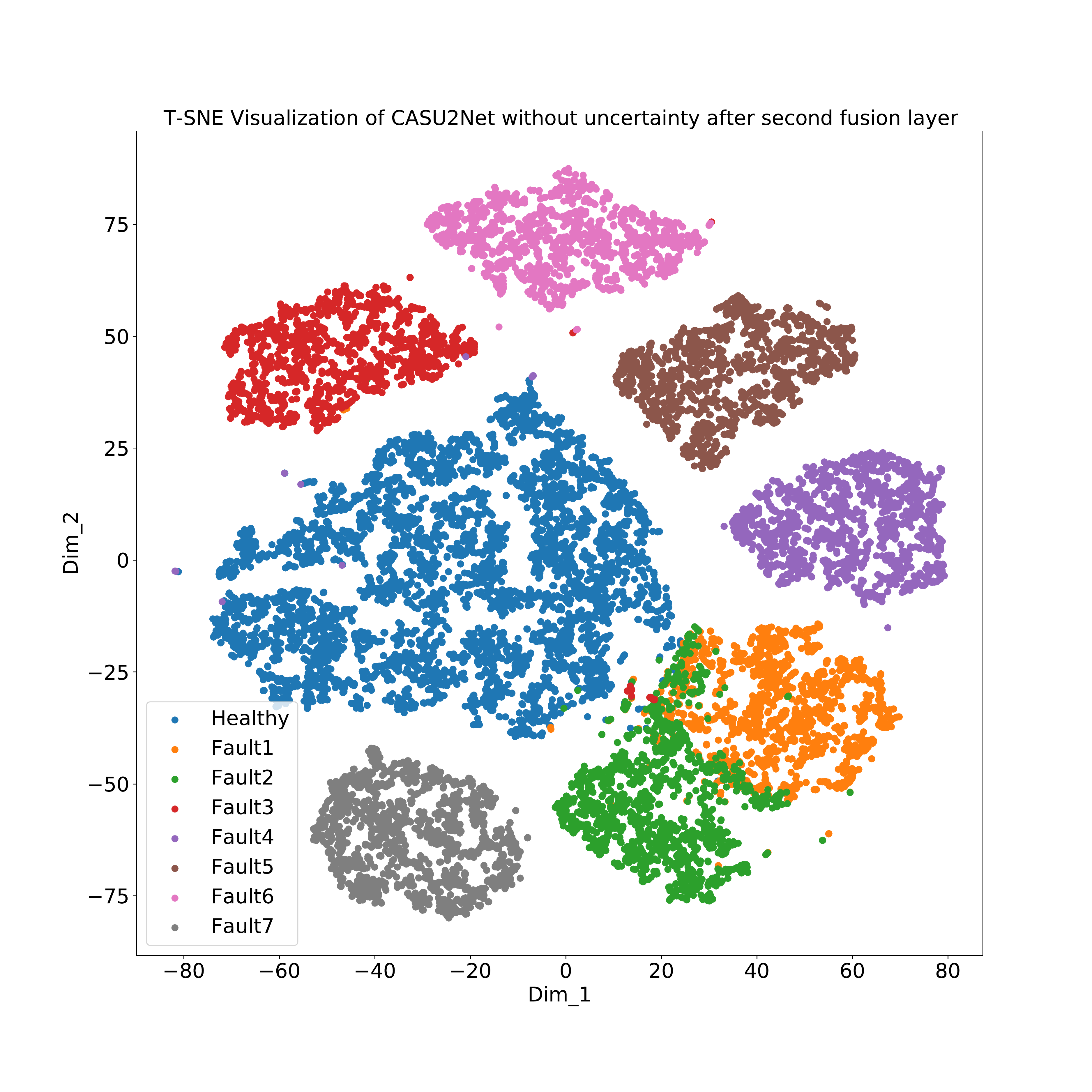}
            \caption{\emph{CASU2Net} second fusion layer without UQ}
            \label{TSNE_CASU2Net_without_second}
    \end{subfigure}
    \caption{T-SNE visualisation of different methods without uncertainty quantification.}
\end{figure*}

\begin{figure*}[h!]
\centering
    \begin{subfigure}[b]{0.3\textwidth}
            \centering
            \includegraphics[width=\textwidth]{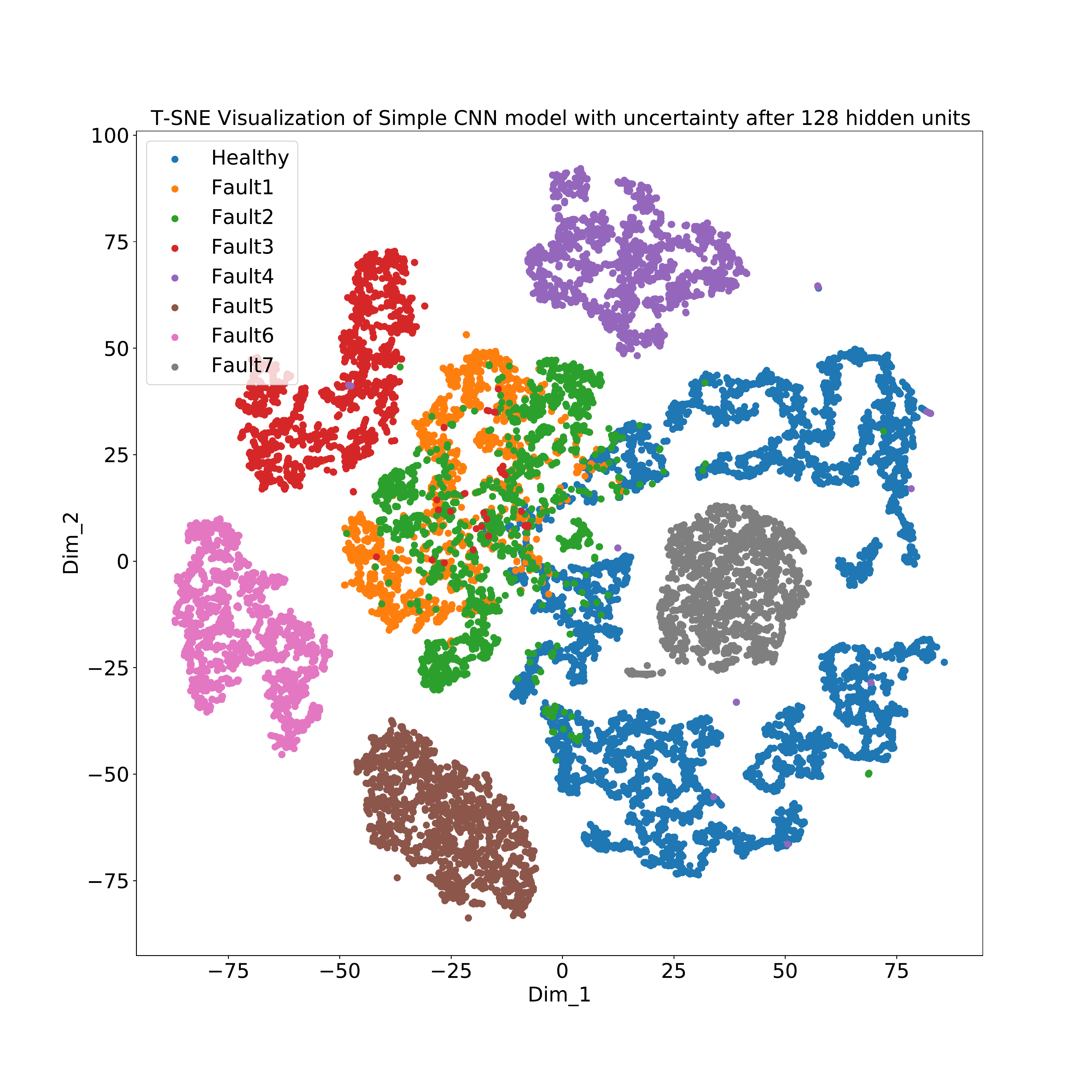}
            \caption{Simple CNN with UQ}
            \label{TSNE_Simple_with}
    \end{subfigure}
    \begin{subfigure}[b]{0.3\textwidth}
            \centering
            \includegraphics[width=\textwidth]{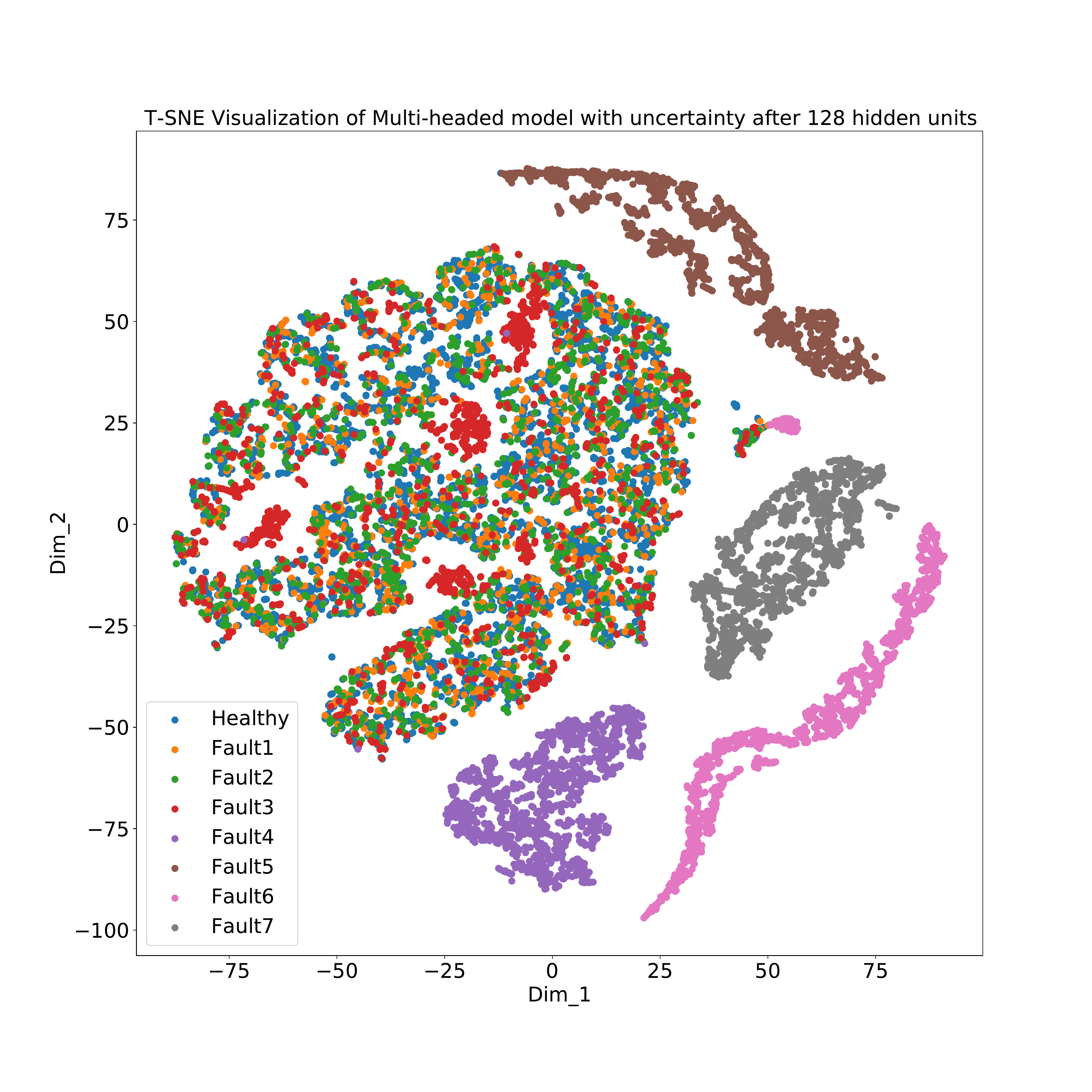} 
            \caption{Multi-headed with UQ}
            \label{TSNE_Multi_with}
    \end{subfigure}  
    \begin{subfigure}[b]{0.3\textwidth}
            \centering
            \includegraphics[width=\textwidth]{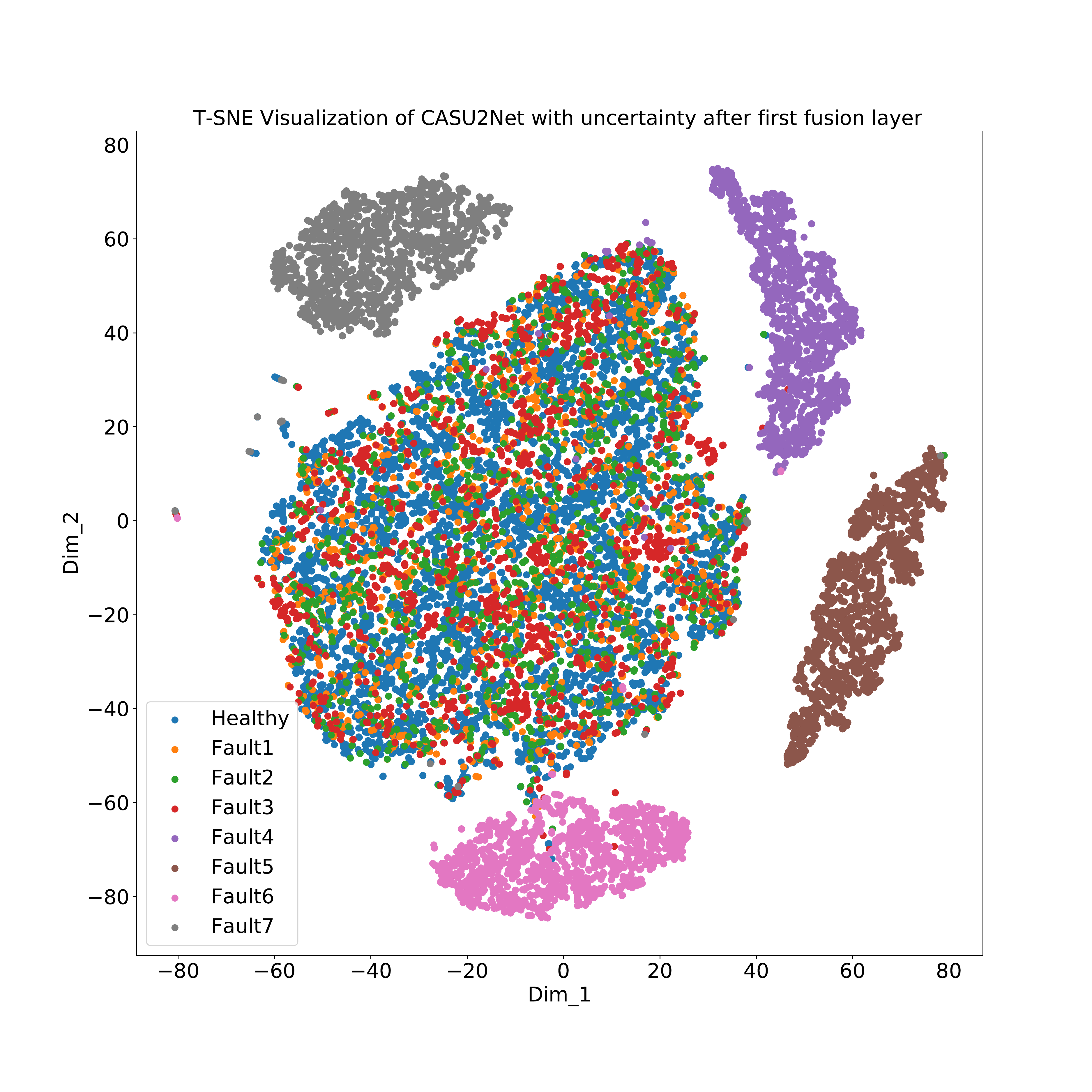}
            \caption{\emph{CASU2Net} first fusion layer with UQ}
            \label{TSNE_CASU2Net_with_first}
    \end{subfigure}
     \begin{subfigure}[b]{0.3\textwidth}
            \centering
            \includegraphics[width=\textwidth]{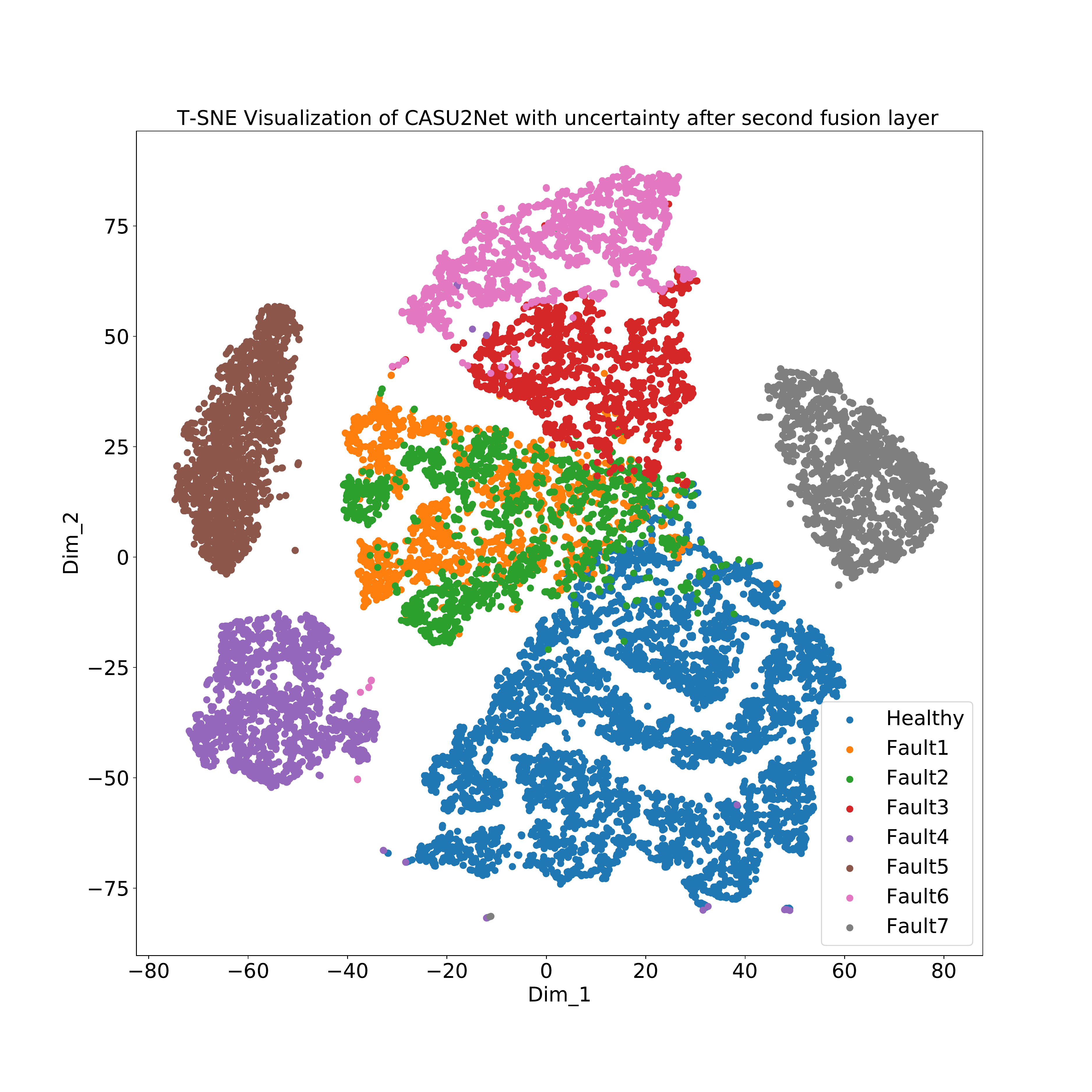}
            \caption{\emph{CASU2Net} second fusion layer with UQ}
            \label{TSNE_CASU2Net_with_second}
    \end{subfigure}
    \caption{T-SNE visualisation of different methods with uncertainty quantification.}
\end{figure*}

In this section, the t-distributed stochastic neighbor embedding (T-SNE) method \cite{van2008visualizing}, which represents a nonlinear dimensionality reduction technique, is employed for layer representation of the proposed models. During the training of the \emph{CASU2Net}, the output of the fusion layers represents the general view of the learned features. This representation illustrates how \emph{CASU2Net} captures time dependencies. As we can see from Figs. \ref{TSNE_CASU2Net_without_first} and \ref{TSNE_CASU2Net_with_first}, after the first fusion layer, faults 4, 5, and 6 become discernible, which denotes that these classes of faults are more comfortable to detect than others. In these Figs, we can see pitch actuator's faults are not distinguished from other classes properly, so we can conclude that pitch actuator's faults represent the most challenging faults. Also, It is shown in Figs. \ref{TSNE_CASU2Net_without_second} and \ref{TSNE_CASU2Net_with_second}
that after the second fusion layer, all classes are separable, and that confirms the results of the \emph{CASU2Net}. On the other hand, by comparing the features after the first and second layers, we can see that by increasing the size of the input signal, the classes become more separable, which confirms the importance of time dependencies for classification. Thus, the \emph{CASU2Net} considers time dependencies well. Moreover, from the T-SNE visualization of the simple CNN and multi-headed model in Figs. \ref{TSNE_Simple_without}, \ref{TSNE_Simple_with}, \ref{TSNE_Multi_without}, and \ref{TSNE_Multi_with}, we can see that these models can not distinguish the pitch actuator's faults well.

As we indicated earlier, the proposed models can predict the test samples well. Although the simple CNN and multi-headed model are not able to detect pitch actuator's faults properly, the \emph{CASU2Net} correctly detects all faults. Also, previous works in the area of fault detection of wind turbines have overlooked uncertainty quantification. In contrast, our designed models have MC dropout as an uncertainty module to assess models' trustworthiness. In \cite{ruiz2018wind}, samples from 13 sensors, each with a length of 600 s is transformed to grayscale images and used for feature extraction. After feature extraction, several classifiers are used for fault classification. It is shown that with 80\% accuracy, the bagged trees provide an improvement over other classifiers \cite{ruiz2018wind}. Compared to the previous approach \cite{ruiz2018wind}, the proposed \emph{CASU2Net} exhibits 99.4\% accuracy, by only using 5 sensors instead of 13. It is also worth mentioning that every one of our samples has a length of only 1.5 s, which is much shorter than the input length reported by \cite{ruiz2018wind}. As a result, our fault detection model can detect faults much faster than the existing models. In other words, when the input length of the fault detection model is more, it needs more time to wait for samples, and hence it acts slower than the other models with shorter input lengths. By using only 5 sensors for fault detection and classification, we can reduce the cost of operation of offshore wind turbines.

\section{Conclusion}
\label{S:7} 

In recent years, different state-of-the-art deep learning methods have demonstrated good performance in various applications. In this paper, to optimally take advantage of the impressive performances of RNN and LSTM, we proposed three different deep learning models, including simple CNN, multi-headed model, and \emph{CASU2Net} (as our main feature fusion model) for fault detection in offshore wind turbines. A review of past studies revealed that none of them quantified uncertainty of the proposed ML and DL models. To fill this gap, we applied MC dropout as a well-known uncertainty quantification technique. The obtained results indicated that our proposed model \emph{CASU2Net} achieves an average accuracy of 99.4\% over all faults, and it outperforms the proposed other models in this study and also the previous models using the same data set. The advantages of our \emph{CASU2Net} are as follows: a) not require any feature extraction, b) contain two-step early fusion layers, c) detects faults faster than the existing models, using the same data set, d) uses smaller number of sensors to reduce the cost of operation of an offshore wind turbine, and finally e) uses uncertainty quantification module to quantify its uncertainty while predictions. Based on the obtained results, we should remark that our data-driven approach can also be used in any time-domain data sets obtained from other sources. Even though the performance of \emph{CASU2Net} is outstanding, there are some suggestions as follows for achieving better results. In the following, we list some of the most important future research directions:

\begin{enumerate}

\item The proposed \emph{CASU2Net} model just takes advantage of the early fusion approach. In contrast, some other fusion techniques (e.g., late fusion as a decision-level fusion) can improve the model's performance. For this reason, we aim to extend the proposed \emph{CASU2Net} model, including at least one late fusion module.     
%\item Model uncertainty plays a significant role in enhancing the trustworthiness of the results obtained by the deep learning algorithms. However, using uncertainty quantification (UQ) methods in deep learning pose some challenges. In this study the UQ methods are not implemented and we believe that proposing a model with UQ methods can improve the performance of current model. Meanwhile, UQ methods can bring the results closer to reality.
%\item 

\item Another issue in our proposed model is the final prediction step. The proposed model has only one prediction step, while by having more predictions and using different models, the performance may improve. To deal with this issue, prediction-based fusion methods can improve the final results. In this regard, a novel weighted ensemble model that combines various deep learning methods is suggested.

\item Moreover, we just simply employ a well-known uncertainty quantification method. However, several new uncertainty quantification methods can also be used. For example, some newest uncertainty quantification methods such as Deterministic Uncertainty Estimation (DUE) \cite{van2021improving}, Mix-n-Match (as an Ensemble approach) \cite{zhang2020mix}, Learned Confidence (LC) \cite{henne2020benchmarking}, etc., can be employed. 

\end{enumerate}

Further recommendations for future research directions are also summarized as follows:
\begin{enumerate}

\item Work with real-world lower frequency data sets.

\item Use some hybrid models with different fusion approaches, either early or late fusion, and also employ ensemble leaning techniques.

\item Consider other faults including the system faults.

\item Reduce even further the number of sensors.

\end{enumerate}

\bibliographystyle{ieeetr}
\bibliography{ref.bib}

\begin{thebibliography}{10}
\bibitem{shanshanoriginal}
Z.~Shanshan, ``Original paper multi-source information fusion technology and
  its engineering application,''
\end{thebibliography}

\end{document}